\pdfoutput=1 

\documentclass[a4paper,11pt]{article}

\usepackage{jheppub} 
\usepackage{amsmath,amssymb}
\usepackage{mathdots}
\usepackage{xcolor}
\usepackage[utf8]{inputenc}
\usepackage{mathtools}
\usepackage[T1]{fontenc}
\usepackage{tikz}
\usepackage{colortbl}
\usepackage{ascmac}
\usepackage{enumitem}

\newcommand{\ud}{\mathrm{d}}
\newcommand{\ub}{\mathrm{b}}
\newcommand{\R}{\mathrm{R}}
\newcommand{\NS}{\mathrm{NS}}

\newcommand{\Tr}{\mathrm{Tr}}
\newcommand{\Teichmuller}{Teichm\"{u}ller }
\newcommand{\Li}{\mathrm{Li}}
\newcommand{\eR}{e_{\rm R}}
\newcommand{\eNS}{e_{\rm NS}}

\DeclareSymbolFont{sfletters}{OML}{cmbrm}{m}{it}
\DeclareMathSymbol{\hvarphi}{\mathord}{sfletters}{"27}
\DeclareMathSymbol{\srho}{\mathord}{sfletters}{"1A}

\newcommand{\SL}{\mathrm{SL}}
\newcommand{\PSL}{\mathrm{PSL}}
\newcommand{\OSp}{\mathrm{OSp}}
\newcommand{\SO}{\mathrm{SO}}
\newcommand{\SU}{\mathrm{SU}}
\newcommand{\Mp}{\mathrm{Mp}}

\def\be{\begin{equation}}
\def\ee{\end{equation}}

\makeatletter
\def\mathcolor#1#{\@mathcolor{#1}}
\def\@mathcolor#1#2#3{%
  \protect\leavevmode
  \begingroup\color#1{#2}#3\endgroup
}
\makeatother

\title{Towards Super \Teichmuller Spin TQFT}

\author[a,b]{Nezhla Aghaei,}
\author[c]{M.K. Pawelkiewicz,}
\author[d]{and Masahito Yamazaki}

\affiliation[a]{Max Planck Institut f\"ur Mathematik, Vivatsgasse 7, 53111 Bonn, Germany.}
\affiliation[b] {Albert Einstein Center for Fundamental Physics, Institute for Theoretical Physics, \\ University of Bern, Sidlerstrasse 5, Bern, ch-3012, Switzerland.}
\affiliation[c]{ Institut de Physique Theorique, CEA Saclay, 91191 Gif Sur Yvette, France.}
\affiliation[d]{Kavli Institute for the Physics and Mathematics of the Universe (WPI), UTIAS, \\ the University of Tokyo, Kashiwa, Chiba 277-8583, Japan.}

\emailAdd{nezhla.aghaee@mpim-bonn.mpg.de}
\emailAdd{mpawelki@ipht.fr}
\emailAdd{masahito.yamazaki@ipmu.jp}

\abstract{The quantization of the \Teichmuller theory has led to the formulation of the so-called \Teichmuller TQFT for 3-manifolds. In this paper we initiate the study of ``supersymmetrization'' of the \Teichmuller TQFT, which we call the super \Teichmuller spin TQFT. We obtain concrete expressions for the partition functions of the super \Teichmuller spin TQFT for a class of spin 3-manifold geometries, by taking advantage of the recent results on the quantization of the super \Teichmuller theory. We then compute the perturbative expansions of the partition functions, to obtain perturbative invariants of spin 3-manifolds. We also comment on the relations of the super \Teichmuller spin TQFT to 3-dimensional Chern-Simons theories with complex gauge groups, and to a class of 3d $\mathcal{N} = 2$ theories arising from the compactifications of the M5-branes.}

\begin{document}

\maketitle

\section{Introduction}

Quantum \Teichmuller theory \cite{Chekhov:1999tn,Kashaev:1998fc} has been a 
fascinating subject with connections to many different topics in mathematics and physics.
While the quantum \Teichmuller theory in itself concerns 2-manifolds, one can use the ingredients from the \Teichmuller theory to define the so-called 
 \emph{\Teichmuller Topological Quantum Field Theory} (\emph{\Teichmuller TQFT}) \cite{Andersen:2011bt,Andersen:2013rxa, Andersen:2018pnw},\footnote{See also 
 \cite{Kashaev:1994pj,MR1848458,Hikami:2007zz,Dimofte:2009yn,Dimofte:2011gm,Dimofte:2012qj,Kashaev:2012cz,Dimofte:2014zga,Garoufalidis:2014ifa}. The topic has also been discussed 
in connection with the 3d--3d correspondence \cite{Terashima:2011qi,Dimofte:2011ju}. Some papers on this topic, with emphasis on the mapping torus geometry, include \cite{Terashima:2011qi,Terashima:2011xe,Dimofte:2011jd,Gang:2012ff,Gang:2015bwa,Gang:2015wya,Chun:2019mal}.} which generates interesting topological invariants of 3-manifolds.

There are several possible extensions of the quantum \Teichmuller theory.~One such generalization is to consider the super \Teichmuller theory,
which studies the moduli space of super Riemann surfaces.
While the super \Teichmuller space has long been playing fundamental roles in 
the perturbative superstring theory,
the systematic analysis of explicit classical coordinate systems \cite{BB,Penner:2015xla}
and the quantization \cite{Aghaei:2015bqi,Aghaei:2019yyb} are
relatively new subjects.\footnote{See also \cite{Ip:2016ojn,Ip:2017msi,Cremonini:2019aao,Bouchard:2019uhx,Huang:2019umm,Stanford:2019vob,Norbury:2020vyi} for some recent papers related to this subject.} 

The goal of the present paper, stimulated by recent developments, 
is to consider the super \Teichmuller counterpart of the 
\Teichmuller TQFT. One of the crucial differences between the super \Teichmuller theory and its
non-supersymmetric (i.e.\ non-super) counterpart is that 
we now need to take into account the dependence on the 
spin structure of the 2-dimensional surface.
Relatedly, the super generalization of the 
\Teichmuller TQFT should depend on the choice of the spin structure of the 3-manifold---we should 
obtain a spin TQFT, not a TQFT. We call this spin TQFT the \emph{super \Teichmuller spin TQFT}.\footnote{Clearly the words ``super'' and ``spin'' are closely related,
and one might be tempted to drop either of them in the terminology. We nevertheless find it useful to emphasize that this is a spin TQFT as defined from the super \Teichmuller theory.}

In this paper we provide algorithms to
compute the partition functions of the super \Teichmuller spin TQFT
for a class of spin 3-manifolds. This leads to concrete integral expressions for the partition functions,
and we study their perturbative expansions.

We also discuss relations with a number of topics from mathematical physics.
It turns out that super \Teichmuller spin TQFT is related to the 3-dimensional Chern-Simons theory 
with complexifications of $\OSp(1|2)$  and $\SO(3)$ gauge groups.
Moreover, we propose to extend the dictionary of the so-called 3d--3d correspondence between supersymmetric 3d gauge theories and 
the 3d Chern-Simons theory \cite{Terashima:2011qi,Dimofte:2011ju};
in our context the partition function of the super \Teichmuller spin TQFT 
is identified with the partition functions of 
3d $\mathcal{N}=2$ supersymmetric theories on the projective space $\mathbb{RP}^3$ \cite{Gang:2019juz,Benini:2011nc}.
This new 3d--3d correspondence originates from compactifications of two
M5-branes on $\mathbb{RP}^3$ times a 3-manifold.

This paper is organized as follows.
We begin in Sec.~\ref{sec:outline} with a summary of our strategy for computing the 
partition function of the super \Teichmuller spin TQFT. 
We first review classical super \Teichmuller theory in Sec.~\ref{sec:classical}.
We then describe the action of the mapping class group in the super \Teichmuller space in Sec.~\ref{sec.new_MCG}.
This result will be uplifted to the quantum super \Teichmuller theory in Sec.~\ref{sec:quantum}.
Based on these results, in Sec.~\ref{sec:Z} we explicitly compute the partition functions of the super \Teichmuller spin TQFT for mapping 
tori associated with the once-punctured torus. 
We further study the relation of the super \Teichmuller spin TQFT with a number of different topics, such as the 3-dimensional
Chern-Simons theory in Sec.~\ref{sec:CS} and the 3d--3d correspondence and M5-branes in Sec.~\ref{sec:3d3d}. 
Finally we suggest possible future problems in Sec.~\ref{sec:summary}.

\section{Outline of Strategy}\label{sec:outline}
Before coming to details, let us outline our strategy. Our construction of the super \Teichmuller spin TQFT is inspired by the 
Atiyah-Segal type axioms in TQFT \cite{MR1001453} (adopted here for a spin TQFT). Recall that a spin TQFT
associates a partition function $Z(M)$ to a closed spin 3-manifold $M$,
and a ``Hilbert space'' $\mathcal{H}(\Sigma)$ to a spin 2-manifold $\Sigma$.
When $M$ is a 3-manifold with boundaries, $Z(M)$ defines an element (a wavefunction)
in the Hilbert space associated with the boundaries of $M$. For example,
if $M$ has two spin 2-manifolds $\Sigma_1$ and $\Sigma_2$ as boundaries, namely
$\partial M=(-\Sigma_1) \cup \Sigma_2$  (where minus here means the orientation reversal),
then we have
\begin {align}
Z(M) \in \mathcal{H}(\Sigma_1)^{*} \otimes \mathcal{H}(\Sigma_2) \simeq \mathrm{Hom}(\mathcal{H}(\Sigma_1), \mathcal{H}(\Sigma_2)) \;.
\end{align}
Similarly, if one has the a 3-manifold $M$ with three 2-manifolds  $\Sigma_{1,2,3}$ (so that
$\partial M=(-\Sigma_1) \cup \Sigma_2\cup \Sigma_3$, then one has
\begin {align}
Z(M) \in \mathcal{H}(\Sigma_1)^{*} \otimes \mathcal{H}(\Sigma_2) \otimes \mathcal{H}(\Sigma_3)  \;. 
\end{align}
This discussion points us to a possible strategy in formulating a 3-dimensional spin TQFT: we first 
start with the formulations of the Hilbert space $\mathcal{H}(\Sigma)$ for a spin 2-manifold $\Sigma$, identify the operators 
acting on their tensor products, and make contact with the geometry of the 3-manifold $M$.
In physics language, this is to adopt the Hamiltonian formulation of the theory. Fortunately, the ``Hilbert space'' $\mathcal{H}(\Sigma)$
associated with a punctured spin 2-manifold $\Sigma$ has already been constructed in the literature
in the context of the quantum super \Teichmuller theory
\cite{Aghaei:2015bqi} and we will  use this as a starting point of our discussion.

The connection between the geometry of the 3-manifold and the boundary 2-manifolds
is particularly pronounced for the mapping torus geometry
\begin{align}
M=(\Sigma\times S^1)_{[\varphi]}\coloneqq (\Sigma\times [0,1])/\sim \;,
\label{mt_def}
\end{align}
where the equivalence class $\sim$ is given by
$(x, 0)\sim (\varphi(x), 1)$ for $x\in \Sigma$, $\varphi \in \mathrm{Aut}(\Sigma)$. This is a non-trivial fibration of $\Sigma$ over $S^1$.
Note that the topology of the mapping torus $(\Sigma\times S^1)_{[\varphi]}$
depends only on the spin mapping class $[\varphi]$, and not on the choice of the automorphism $\varphi$ within the class $[\varphi]$. While the mapping torus \eqref{mt_def} in itself does not have  a boundary,
we can ``cut open'' the mapping torus into a mapping cylinder $M=(\Sigma\times [0,1])_{\varphi}$.
This is a 3-manifold with two boundaries $\partial M=(-\Sigma) \cup \Sigma$,
where the boundary conditions at the two surfaces $\Sigma$ are twisted by $\varphi$.
The spin TQFT then associates an operator 
\begin {align}
\hvarphi=Z\left[(\Sigma\times [0,1])_{\varphi}\right] \in  \mathrm{End}(\mathcal{H}(\Sigma)) \;.
\end{align}

Once we identify the operator $\hvarphi$,
one can then compute the spin 3-manifold invariant
$Z[(\Sigma\times S^1)_{[\varphi]}]$ by a suitable trace:\footnote{More precisely we need to insert suitable projection operators into the trace to fix a 3d spin structure, as we will discuss in Sec.~\ref{subsec:def_Z}.}
\begin{align}
Z\left[(\Sigma\times S^1)_{[\varphi]}\right]=\textrm{Tr}(\hvarphi) \;.
\end{align}

Our discussion naturally generalizes similar discussions for the 
non-supersymmetric quantum \Teichmuller theory \cite{Terashima:2011qi,Terashima:2011xe,Terashima:2013fg,Gang:2015bwa} to supersymmetric settings.
We will find, however, that there are important differences in the formulations of the theory,
both technically and conceptually. One of the crucial differences is that the
super \Teichmuller TQFT is a spin TQFT (not a TQFT), and hence depends on the choice of the 
spin structure of the 3-manifold (and hence of the 2-manifold). In order to explain this point, let us first begin in the next section 
with the summary of the classical super \Teichmuller theory.

\section{Review: Classical Super \Teichmuller Theory}\label{sec:classical}
In this section we briefly summarize the essence of the super \Teichmuller theory (see e.g.\ \cite{BB} for details).
We first introduce coordinate system for the super \Teichmuller space in Sec.~\ref{3.1}, whose definition requires 
combinatorial spin structures in Sec.~\ref{3.2}. We also discuss coordinate transformations in Sec.~\ref{3.3}. 

\subsection{Generalities on Super \Teichmuller Space}\label{3.1}

A super Riemann surface $\Sigma_{g,n}$ is a 1-dimensional complex supermanifold with genus $g$ and the number of punctures $n$. For our goals it will be most convenient to simply define super Riemann surfaces as quotients of the super upper half-plane by suitable discrete subgroups $\Gamma$ of $\OSp(1|2)$.

A natural map from  $\OSp(1|2)$ to $\SL(2,\mathbb{R})$ may be defined by 
mapping the odd generators to zero. The image of 
$g\in \OSp(1|2)$  under this map will be denoted as $g^{\sharp}\in \SL(2,\mathbb{R})$. A discrete subgroup of $\Gamma$ of $\OSp(1|2)$ 
such that $\Gamma^\sharp$ is a Fuchsian group is called a super Fuchsian group. 
In fact, a super Fuchsian group is a finitely generated discrete subgroup of $\OSp(1|2)$ which reduces to a Fuchsian group.

The super upper half-plane is defined as $\mathbb{H}^{1|1} \coloneqq \left\{ (x,\theta)\in \mathbb{C}^{1|1}:{\rm Im}(x)>0 \right\} $ and a super Riemann surface of constant negative curvature will be defined as 
a quotient of the super upper half-plane $\mathbb{H}^{1|1}$ by a super Fuchsian group $\Gamma$,
\begin{equation}
\Sigma_{g,n} \coloneqq \mathbb{H}^{1|1}\slash \Gamma \;.
\end{equation}
The group $\OSp(1|2)$ is the group of automorphisms of $\mathbb{H}^{1|1}$ under which the metric is invariant.

We can define the super \Teichmuller space $\mathcal{ST}_{g,n}$ of super Riemann surfaces $\Sigma_{g,n}$ of genus $g$ with $n$ punctures as the quotient \cite{MR943988,MR923633,MR1222942}
\begin{equation}
\mathcal{ST}_{g,n} = \big\{ \rho: \pi_1(\Sigma_{g,n}) \to \OSp(1|2) \big\}\, \slash \, \OSp(1|2) \;,
\end{equation}
where $\rho$ is a discrete 
representation of fundamental group $\pi_1(\Sigma_{g,n})$ into $\OSp(1|2)$ whose image is super Fuchsian. There is always a non-supersymmetric Riemann surface $\Sigma_{g,n}^\sharp$ associated to each super Riemann
surface, defined as the quotient of the upper half-plane $\mathbb{H}$ by $\Gamma^\sharp$. 
Notions such as ideal triangulations, where the vertices of the triangulations are on the boundaries of the super upper half-plane, will therefore have obvious counterparts in the theory of super Riemann surfaces. 

The isometry group $\OSp(1|2)$ acts on the super upper half-plane $\mathbb{H}^{1|1}$ by generalized M\"obius transformations of the form
\begin{align}
&x \longrightarrow x'=\frac{ax+b+\gamma \theta}{cx+d+\delta\theta} \;, \\
&\theta \longrightarrow \theta'=\frac{\alpha x+\beta+e \theta}{cx+d+\delta\theta}\,.
\end{align}

We can define two types of invariants, even $Z$ and odd $\xi$, under the generalized M\"obius transformations~\cite{MR1095783}.
The first is the super conformal cross-ratio (even super Fock coordinate) $Z$ defined for four points $P_i=({x_i | \theta_i})$, $i=1,\dots,4$ in the super upper half-plane:
\begin{equation}\label{conformalinvariantt}
Z\coloneqq \frac{X_{14}X_{23}} {X_{12}X_{34}}\;,
\end{equation}
where $X_{ij} \coloneqq x_i-x_j-\theta_i\theta_j$. 
This is a natural generalization of the cross-ratio (Fock coordinate) in the non-supersymmetric \Teichmuller theory.
The second is an odd invariant $\pm \xi$ 
associated to a collection of three
points $P_i=({x_i | \theta_i})$, $i=1,2,3$:
\begin{equation}\label{oddinv}
 \xi \coloneqq \pm \frac{x_{23}\theta_1+x_{31}\theta_2+x_{12}\theta_3 - \frac{1}{2}\theta_1\theta_2\theta_3}{(X_{12}X_{23}X_{31})^\frac{1}{2}} \;,
\end{equation}
where $x_{ij} \coloneqq x_i-x_j$. Note that at this point we have not
fixed the sign ambiguity.\footnote{For this reason this odd invariant was called the pseudo-invariant in \cite{MR1095783}.}

\subsection{Combinatorial Spin Structures}\label{3.2}

In this part we discuss combinatorial spin structures. We define the Kasteleyn orientation and explain its relation with the spin structure. 

\subsubsection{Kasteleyn Orientation}\label{3.2.1}

Suppose that we choose an ideal triangulation for the super Riemann surface.
We can then introduce a collection of coordinates $Z$ (for each edge) and $\xi$ (for each face)  
as in \eqref{conformalinvariantt} and \eqref{oddinv}---it is known that these coordinates provide a good coordinate system for the super \Teichmuller space \cite[Theorem 4.3.10]{BB}.

For this purpose, however, we still need an extra data to fix the signs in the definition of the odd invariant $\xi$ \eqref{oddinv}.
This extra data allows us to define the
lifts of the punctures $P_i\in\mathbb{P}^{1|1}$ to 
points $\hat{P}_i$ on the double cover $\mathbb{S}^{1|1}$ over $\mathbb{P}^{1|1}$. Note that the 
even part of $\mathbb{P}^{1|1}$ is the real projective line $\mathbb{R}\mathbb{P}^1$ with group of automorphisms
$\PSL(2,\mathbb{R})$, while the even part of $\mathbb{S}^{1|1}$ is a double cover of $\mathbb{R}\mathbb{P}^1$
with group of automorphisms $\SL(2,\mathbb{R})$. Lifting the vertices of the triangulation of $\mathbb{H}^{1|1}$ 
to $\mathbb{S}^{1|1}$ should therefore be accompanied with a lift of the Fuchsian group $\Gamma^\sharp\subset \PSL(2,\mathbb{R})$
to a subgroup of $\SL(2,\mathbb{R})$.  It is known that the definition of such a lift depends on the choice of 
the spin structure on $\Sigma$ \cite{MR2075914}, and indeed different connected components of the super \Teichmuller space are indexed by such spin structures. 
Therefore, we need to fix a combinatorial spin structure on the ideal triangulation.
This is achieved by the so-called Kasteleyn orientation \cite{Kasteleyn,MR2335773,MR2410902},
which we now explain.

\begin{figure}[htbp]
	\centering
	\includegraphics[width=0.4\textwidth]{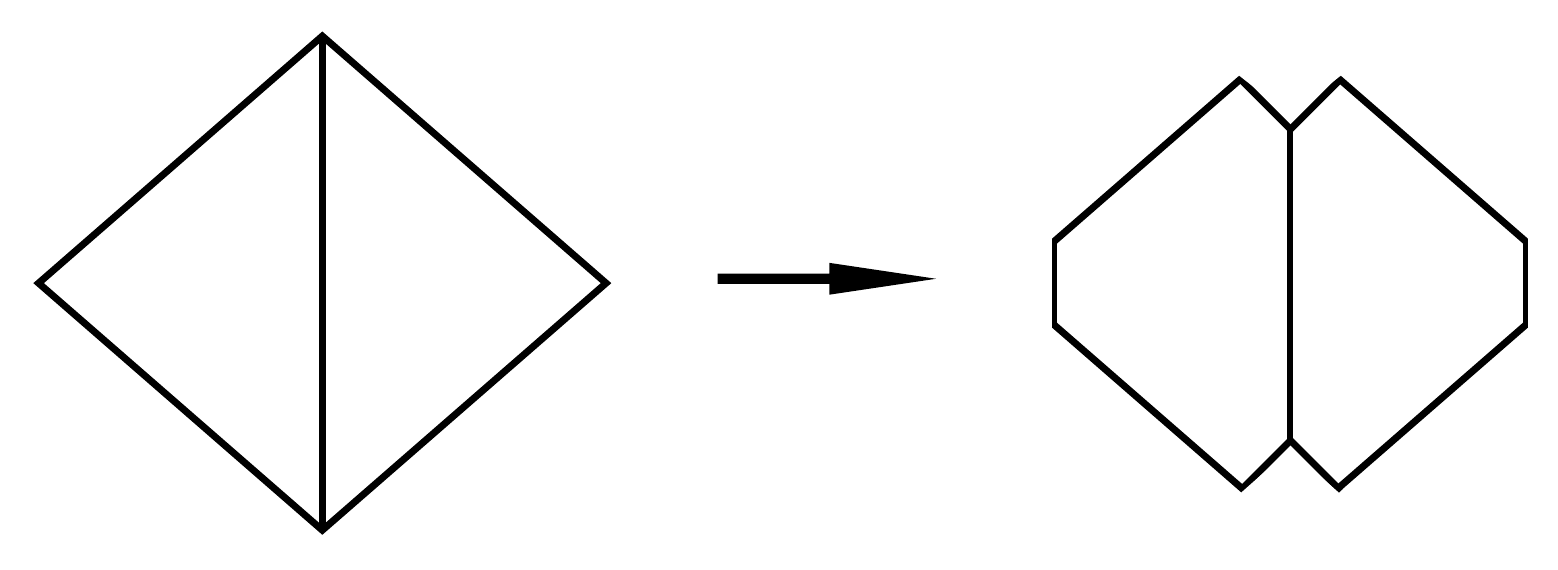}
	\caption{The hexagonalization of an ideal triangulation.}
	\label{dischexagon}
\end{figure}

For each given ideal triangulation we consider the canonical orientation induced from that of the 2-dimensional surface. 
One can also define the  hexagonalization by ``cutting the corners of triangles'' as in Fig.~\ref{dischexagon}.
The Kasteleyn orientation is an orientation of the boundary edges of the resulting hexagons
such that for every face of the resulting graph the number of edges oriented against the orientation of the surface is odd. 

\begin{figure}[htbp]
	\centering
	\includegraphics[width=0.4\textwidth]{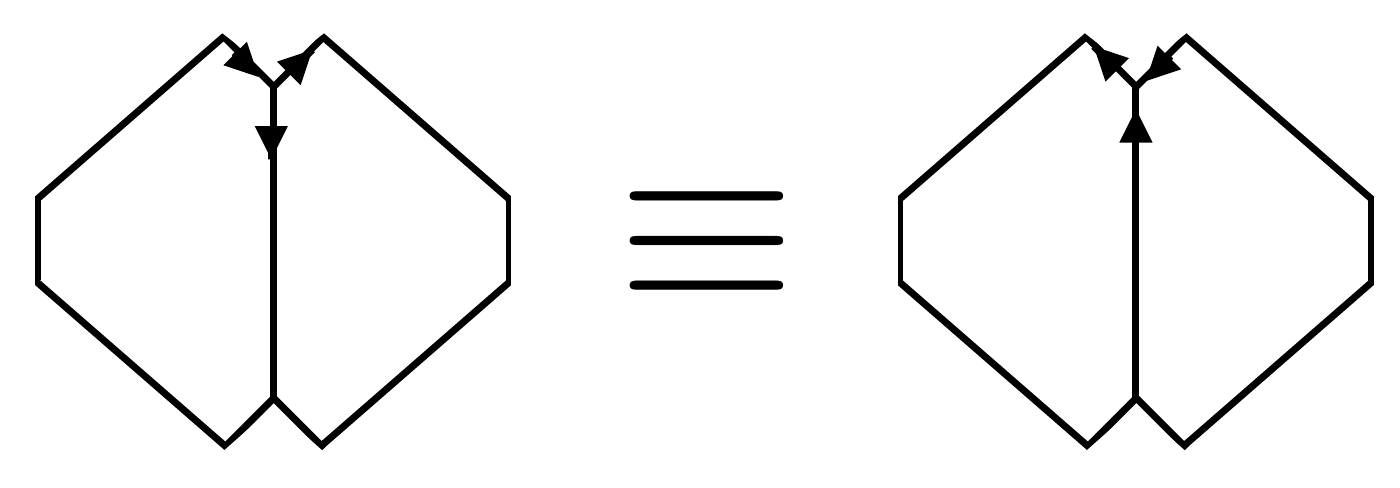}
	\caption{Equivalence between two Kasteleyn orientations.}
	\label{equalkastlegs}
\end{figure}

Two Kasteleyn orientations are said to be equivalent if they are
related by the reversal of orientations of all the edges meeting at 
the same vertex, as illustrated in Fig.~\ref{equalkastlegs}. 

The results of Cimasoni and Reshetikhin \cite{MR2335773,MR2410902} say that the 
equivalence class of the Kasteleyn orientation is in one-to-one correspondence
with the spin structure.

\begin{figure}[htbp]
	\centering
	\includegraphics[width=0.45\textwidth]{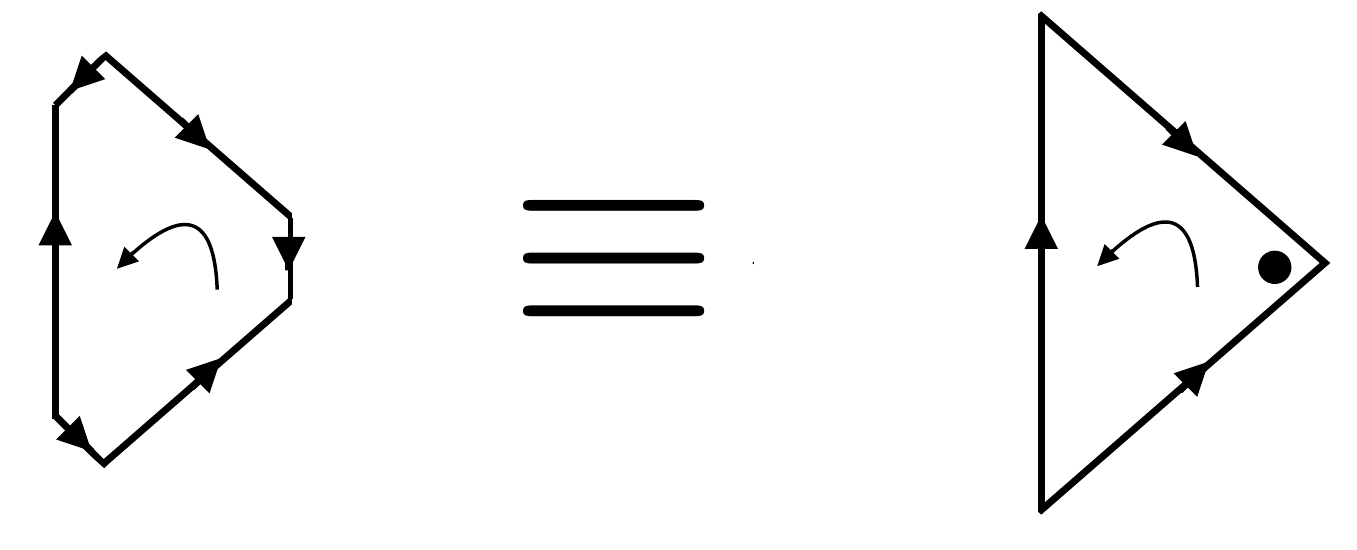}
	\caption{Equivalence between a Kasteleyn orientation on a hexagon and that on a dotted triangle.}
	\label{kasteleyn2}
\end{figure}

For practical applications, it is often cumbersome to represent a Kasteleyn 
orientation with hexagonalization.
One can instead 
introduce a dotted notation on the triangles:
if any corners of the triangulation are dotted (or undotted), it means that the edge of the hexagon associate to that corner
 has the opposite (or the same) orientation from the surface orientation.
An illustration of this procedure is given 
in Fig.~\ref{kasteleyn2} (the surface orientation in these figures is counterclockwise).

\subsubsection{Example: Once-Punctured Torus}\label{3.2.2}

For illustration let us discuss spin structures of once-punctured torus $\Sigma _{1,1}$ in detail.

\begin{figure}[htbp] 	
	\centering
	\includegraphics[width=0.9\textwidth]{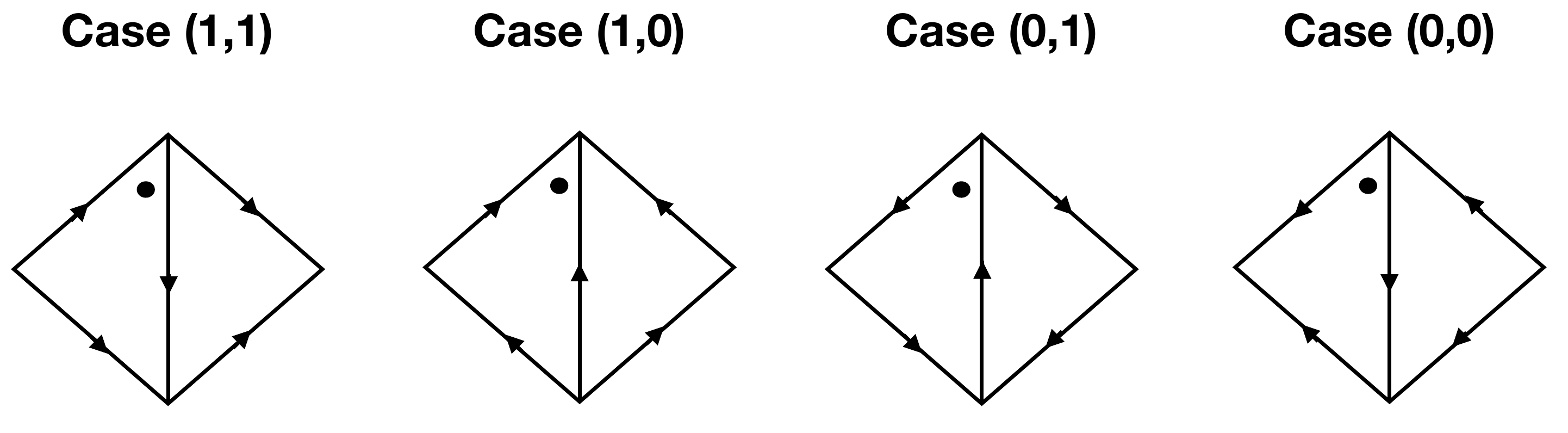}
	\caption{Four different spin structures on the once-punctured torus, as represented by (equivalence classes of) four dotted Kasteleyn orientations.}
	\label{spins}
\end{figure}

The once-punctured torus can be triangulated by two ideal triangles,
and by working out the combinatorics we find that there exist four possible equivalence classes of Kasteleyn orientations as it is
shown in Fig.~\ref{spins}.

Having four equivalence classes of Kasteleyn orientations is  consistent with the fact that there exist four spin structures on the once-puncture torus.
It is known that the space of spin structures is equivalent with $H^1(\Sigma, \mathbb{Z}_2)$ as an 
affine space (i.e.\ if we fix a base point).\footnote{A spin structure can be
identified with a quadratic form on  $H^1(\Sigma, \mathbb{Z}_2)$ \cite{MR588283}.}
For the once-puncture torus this $\mathbb{Z}_2$ cohomology is simply given 
by $H_1(\Sigma_{1,1}, \mathbb{Z}_2)=\mathbb{Z}_2\oplus \mathbb{Z}_2$,
where each $\mathbb{Z}_2$ is associated with the  $\alpha$- and $\beta$-cycles of the torus.\footnote{
In general we have a choice of either Ramond or Neveu-Schwarz boundary condition around the 
puncture. The puncture, however, is always a Neveu-Schwarz puncture for the once-punctured torus,
where $(-1)^{\sigma}=+1$ around the puncture.
This is because the monodromy around the puncture is given by $\alpha\beta \alpha^{-1} \beta^{-1}$
inside the fundamental group, which trivializes in the $\mathbb{Z}_2$ cohomology.}
In Fig.~\ref{spins} we have already shown the corresponding values of $\mathbb{Z}_2\oplus \mathbb{Z}_2$,
which can be derived by the rules explained in Appendix \ref{appendixA}.

\subsection{Coordinate Transformations}\label{3.3}

In Sec.~\ref{3.2.1} we explained that a Kasteleyn orientation can be 
used to fix the signs of the odd invariants. We then have a well-defined 
coordinate system for the super \Teichmuller space for a given dotted triangulation,
i.e.\ for a given Kasteleyn orientation of an ideal triangulation.

However, there is no unique choice of ideal triangulation for a given super Riemann surface. 
In addition, one spin structure corresponds to multiple Kasteleyn orientations inside an equivalence class. It is thus necessary to determine how the coordinates transform under the changes of 
the ideal triangulations and Kasteleyn orientations \cite{BB}.

\paragraph{Push Out}
Let us consider a  move describing a change of Kasteleyn orientation which leaves the spin structure unchanged. 

\begin{figure}[htbp]
	\centering
	\includegraphics[width=0.5\textwidth]{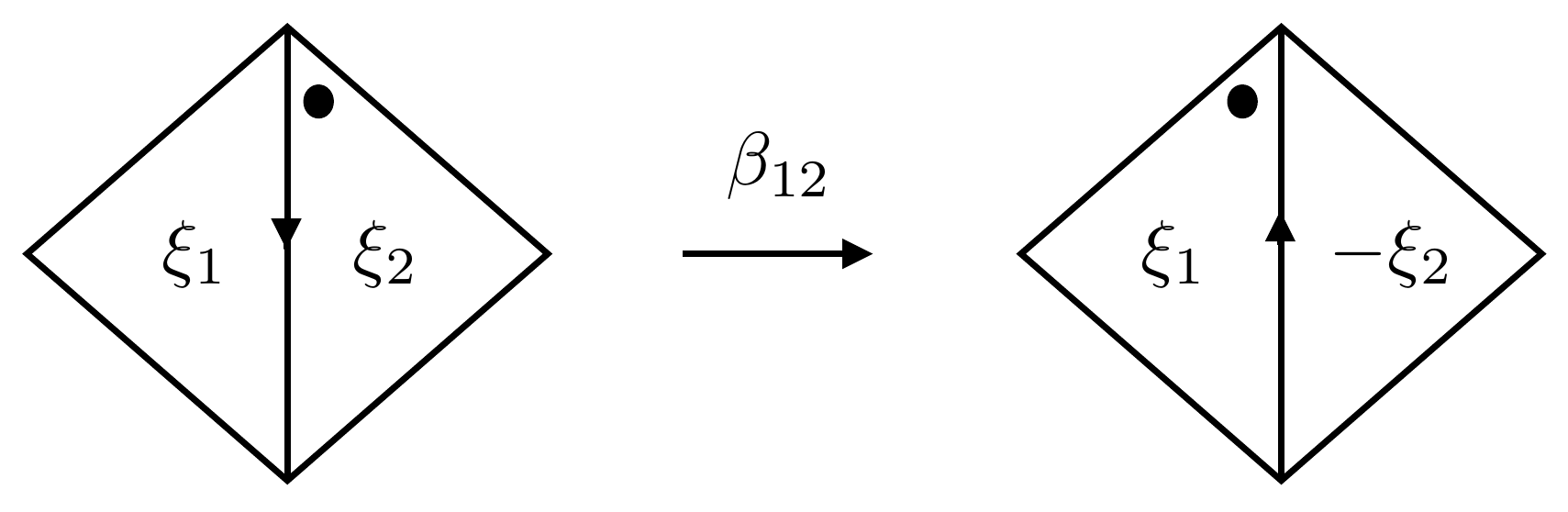}
	\caption{The pictorial representation of a (left) push out on triangles with one dot.}
	\label{pushleftdot}
\end{figure}

In terms of dotted triangles, one can pictorially represent this 
by moving a dot from  from one dotted triangle to another, as in  Fig.~\ref{pushleftdot}---we call this operation
a ``push out." In the figure we have shown the action on the odd invariants:  the invariant of the left hexagon stays the same, while the invariant of the right changes sign.

We can moreover define an inverse of a (left) push out, which we will call a right push out.  On the odd invariants, it acts in the same way as the left push out.

\begin{figure}[htbp] 
	\centering
	\includegraphics[width=0.5\textwidth]{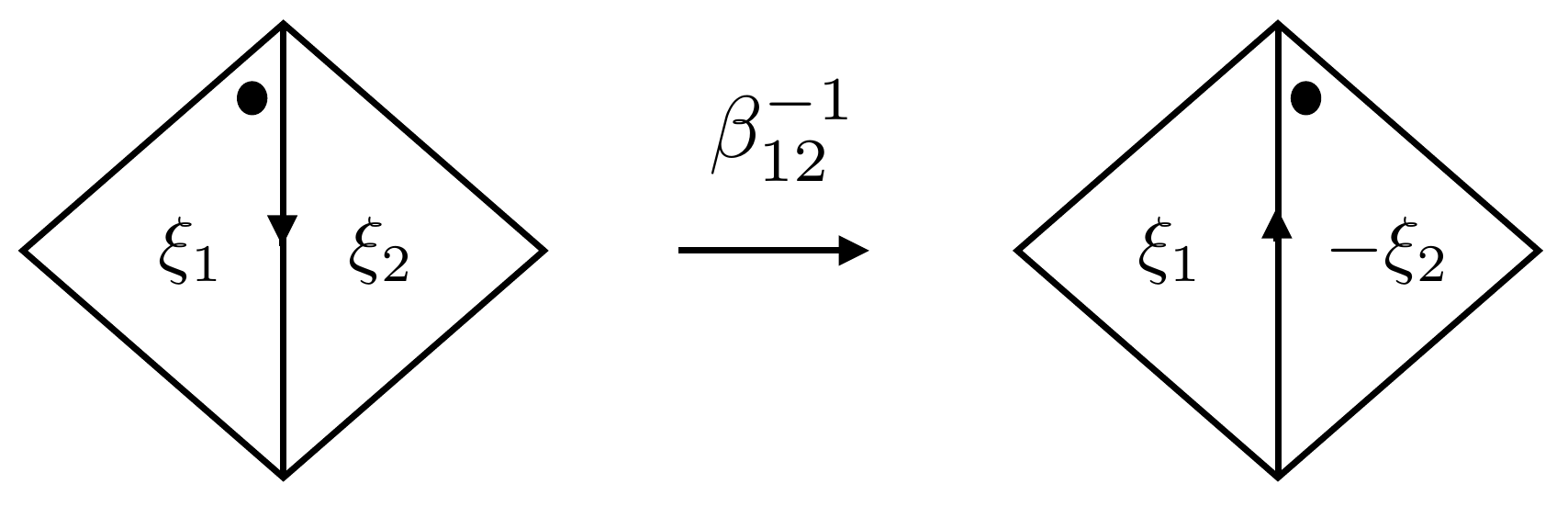}
	\caption{The pictorial representation of a right push out on triangles with one dot. }
	\label{pushrightdot}
\end{figure}

\begin{figure}[h] 
	\centering
	\includegraphics[width=0.55\textwidth]{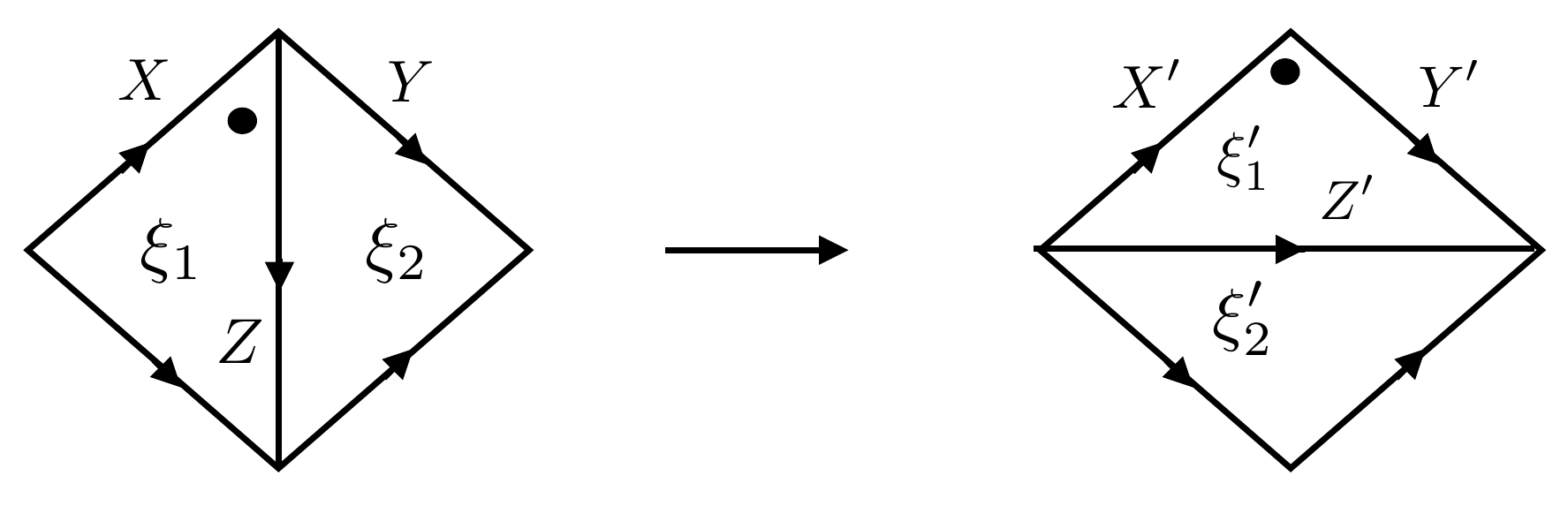}
	\caption{A superflip changes the diagonal of a quadrilateral.}
	\label{flip_show}
\end{figure}

\paragraph{Superflip}
Let us next discuss the change of ideal triangulations.
This can be achieved by a superflip operation as in Fig.~\ref{flip_show},
where we change the the diagonal in a quadrilateral.

Note that in the superflip we have simultaneously changed the Kasteleyn orientation
so that we keep the spin structure. The superflip of Fig.~\ref{flip_show}
generates the coordinate transformation \cite[Figure 5.3]{BB}:
\begin{align}\label{flip_formula}
\left(\begin{array}{l}
X \\ Y\\  Z\\ \xi_1 \\ \xi_2
\end{array}\right)
\to \left(\begin{array}{l}
X'\\ Y'\\  Z'\\ \xi_1' \\ \xi_2'
\end{array}\right)
=
\left(
\begin{array}{l}
 X (1+ Z+ \xi_1\xi_2 \sqrt{Z}) \\
 Y (1 + Z^{-1} + \xi_1\xi_2 \sqrt{Z^{-1}})^{-1}  \\
 Z^{-1}\\
 (1+Z)^{-1/2} (\xi_1-\xi_2\sqrt{Z})\\
 (1+Z)^{-1/2} (\xi_1\sqrt{Z}+\xi_2)
\end{array}
\right) \;.
\end{align}

The coordinate transformations depend crucially on the choice of the Kasteleyn orientations.
In general one can derive the transformation formulas using 
the results of \cite[Chapter 5]{BB} in combination of suitable push outs.  Fig.~\ref{flip_show} is only one of the four equivalence classes of 
Kasteleyn orientations for the quadrilateral.

\section{Mapping Class Group}\label{sec.new_MCG}

In this section we explicitly work out the effects of the mapping class group
on the coordinates of the super \Teichmuller space introduced previously.
While our formalism works in general, we will discuss the example of the once-punctured torus $\Sigma _{1,1}$ in detail.
We begin in Sec.~\ref{sec.MCG} with general reminder on the mapping class group and its action on spin structures.
We then discuss the mapping class group actions on odd and even structures in Sec.~\ref{sec.odd} and Sec.~\ref{sec.even} respectively.

\subsection{Mapping Class Group Generalities}\label{sec.MCG}

Let us discuss the mapping class group for the once-punctured torus. 
In contrast with the flips and push outs, this in general changes the spin structure.

The mapping class group for the once-punctured torus is $\SL(2, \mathbb{Z})$.\footnote{The orientation-preserving subgroup is $\PSL(2, \mathbb{Z})$. We will later find that with fermions we need to consider the double cover of $\SL(2, \mathbb{Z})$, the metaplectic group $\Mp(2, \mathbb{Z})$.}
This group acts on $\alpha$- and $\beta$-cycles, which are generators of $H^1(\Sigma, \mathbb{Z})$, as
\begin{align}
\left(\begin{array}{c}
\alpha \\
\beta
\end{array}\right)
\mapsto 
\left(\begin{array}{cc}
a&b \\
c&d
\end{array}\right)
\left(\begin{array}{c}
\alpha \\
\beta
\end{array}\right)
\;,
\quad
\left(\begin{array}{cc}
a&b \\
c&d
\end{array}\right)
\in \SL(2, \mathbb{Z}) \;.
\end{align}
We choose the generators of the mapping class group to be
\begin{align}
L=\begin{pmatrix}
 1 & 1\\
 0 & 1
\end{pmatrix}, ~~~~
R=\begin{pmatrix}
1 & 0\\
1 & 1
\end{pmatrix},
\end{align}
representing Dehn twists along $\alpha$ and $\beta$-cycles. More concretely, these flips change the $\alpha$ and $\beta$-cycles of the torus as
\begin{align} 
&L:\alpha\rightarrow \alpha+\beta \;,~~\beta \rightarrow \beta \;,
&& R:\alpha \rightarrow  \alpha \;,~~\beta \rightarrow \alpha+\beta \;.
\end{align}

Let us choose a spin structure  $\sigma=(\sigma(\alpha), \sigma(\beta))\in \mathbb{Z}_2\oplus \mathbb{Z}_2$ on the 2-manifold $\Sigma$
and consider the action of the mapping class group element. This is known to be an affine transformation (cf.\ \cite{MR860317})
\begin{align}
\left(\begin{array}{c}
\sigma(\alpha)\\
\sigma(\beta)
\end{array}\right)
\mapsto 
\left(\begin{array}{cc}
a&b \\
c&d
\end{array}\right)
\cdot
\left(\begin{array}{c}
\sigma(\alpha)\\
\sigma(\beta)
\end{array}\right)
+
\left(\begin{array}{c}
ab \\
cd
\end{array}\right) \;.
\label{aff}
\end{align}
We have two orbits under the mapping class group:
an even orbit $\sigma=(0,0), (0,1), (1,0)$ and an isolated odd orbit $\sigma=(1,1)$ (see the left figure of Fig.~\ref{map_new}).


We can choose a different set of the generators for the  $\SL(2, \mathbb{Z})$
mapping class group. For example, we can use the $S$ and $T$ generators
\begin{align}\label{ST}
& S = L^{-1} R L^{-1}=
\left(\begin{array}{cc} 0 & -1\\ 1 & 0 \end{array}\right)  \;, 
\quad T = L=\left(\begin{array}{cc} 1 & 1\\ 0 & 1 \end{array}\right) \;.
\end{align}
The change of the spin structure under these generators is shown in the right figure of Fig.~\ref{map_new}.

\subsection{Mapping Class Group: Odd Spin Structure}\label{sec.odd}

We wish to describe mapping class actions inside the framework of the quantum \Teichmuller theory.
The basic idea is simple: starting with a dotted ideal triangulation,
we apply the mapping class group action, namely to change the fundamental region of the torus.
The result will be another dotted ideal triangulation, to which we can associate another coordinate chart of the 
super \Teichmuller space.

While such an operation in general changes the spin structure, let us
here consider the odd spin structure, namely the type $(1,1)$ spin structure,
so that we are back to the same spin structure (and hence of the same connected component of the 
super \Teichmuller space).

\begin{figure}[htbp]
	\centering
	\includegraphics[width=1\textwidth]{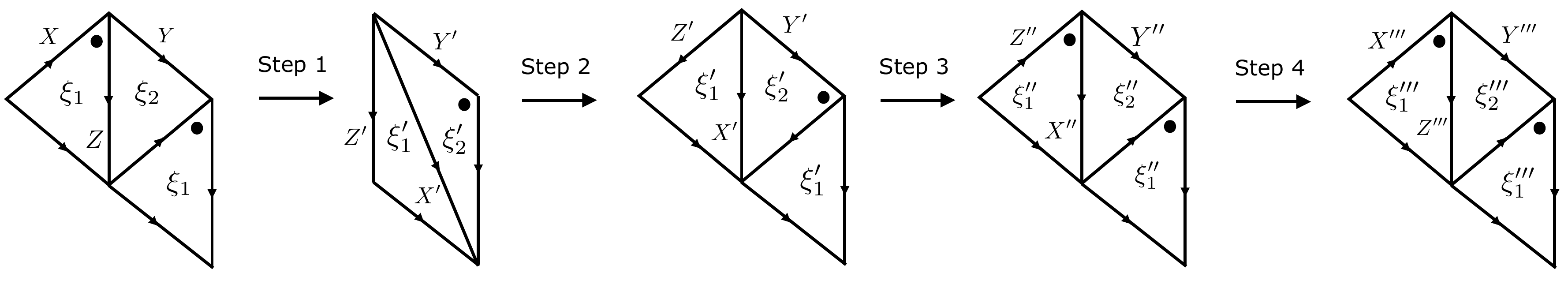}
	\caption{The action of $L_{(1)}$ on a dotted triangulation of type $(1,1)$.}
	\label{L1}
\end{figure}

The steps for deriving the action of the $L$ generator are shown in Figs.~\ref{L1}.
Notice that in writing down the expression for the $L$ generator we
need to make sure that we come back to the same coordinate chart of the 
super \Teichmuller space,
and this requires suitable superflips and push outs from Sec.~\ref{3.3}.

In the leftmost figure of Fig.~\ref{L1},
we denote the odd variable of the dotted triangle by $\xi_1$, and of the undotted triangle by $\xi_2$.
The first step is the superflip in the edge $X$ and we have the following transformations of the super Fock coordinates\footnote{This formula can be derived by the results of \cite[Chapter 5]{BB},
and is a variant of the formula \eqref{flip_formula}. 
Notice that compared with our formula \eqref{flip_formula} for a general surface
we here have a power of $\pm 2$ for the factors $1 + X^{\pm 1}+ \xi_1\xi_2 \sqrt{X}^{\pm 1}$. This is a special feature of the once-punctured torus.}:
\begin{align}
\begin{split}
X'&= X^{-1} \;, \\
Y' &= Y (1 + X + \xi_1\xi_2 \sqrt{X})^2 \;,\\
Z' &= Z (1+ X^{-1} + \xi_1\xi_2 \sqrt{X^{-1}})^{-2}\;, \\
\xi_1' &= (1+X)^{-1/2} (-\xi_1+\xi_2\sqrt{X})\;,  \\
\xi_2' &= (1+X)^{-1/2} (\xi_1\sqrt{X}+\xi_2)\;.
\end{split}
\end{align}

The second step is an $SL(2,\mathbb{Z})$ transformation. As shown in Fig.~\ref{L1}, the only effect of this 
is to change the fundamental region of the torus according to the $L$ action,
and we preserve all the coordinates associated to edges and faces. This nevertheless is 
a rather crucial step for the mapping class group action.

In the third step, we go back to the original dotted triangulation by  a push out, 
leading to the transformation
\begin{align}
&X'' = X' \;,~~~Y'' = Y'\;,~~~Z'' = Z'\;,~~~\xi_1'' = -\xi_1'\;,~~~\xi_2'' = \xi_2'\;.
\end{align}

We now are back in the same coordinate chart, except in the last step we need to exchange $X$ and $Z$ variables:
\begin{align}
&X''' = Z'' \;, 
~~~Y''' = Y''\;, 
~~~Z''' = X''\;, 
~~~\xi_1''' = \xi_1''\;, 
~~~\xi_2'''= \xi_2'' \;.
\end{align}
Therefore, considering the composition of all those steps one has
\begin{align}
\begin{split}
X''' &= Z (1+ X^{-1} + \xi_1\xi_2 \sqrt{X^{-1}})^{-2}\;, \\
Y''' &= Y (1 + X + \xi_1\xi_2 \sqrt{X})^2 \;, \\
Z''' &= X^{-1}\;, \\
\xi_1''' &= (1+X)^{-1/2} (\xi_1-\xi_2\sqrt{X})\;, \\
\xi_2''' &= (1+X)^{-1/2} (\xi_1\sqrt{X}+\xi_2) \;.
\end{split}
\end{align}

Let us summarize this result as a coordination transformation $L_{(1,1)}$, the $L$-generator action on the
super \Teichmuller space equipped with the
type $(1,1)$ spin structure:
\begin{align}\label{L1_formula}
\begin{split}
L_{(1,1)}\colon \qquad 
\left(\begin{array}{l}
X \\ Y\\  Z\\ \xi_1 \\ \xi_2
\end{array}\right)
\to \left(
\begin{array}{l}
 Z (1+ X^{-1} + \xi_1\xi_2 \sqrt{X^{-1}})^{-2} \\
 Y (1 + X + \xi_1\xi_2 \sqrt{X})^2  \\
 X^{-1}\\
 (1+X)^{-1/2} (\xi_1-\xi_2\sqrt{X})\\
 (1+X)^{-1/2} (\xi_1\sqrt{X}+\xi_2)
\end{array}
\right) \;. 
\end{split}
\end{align}

\begin{figure}[htbp]
	\centering
	\includegraphics[width=1\textwidth]{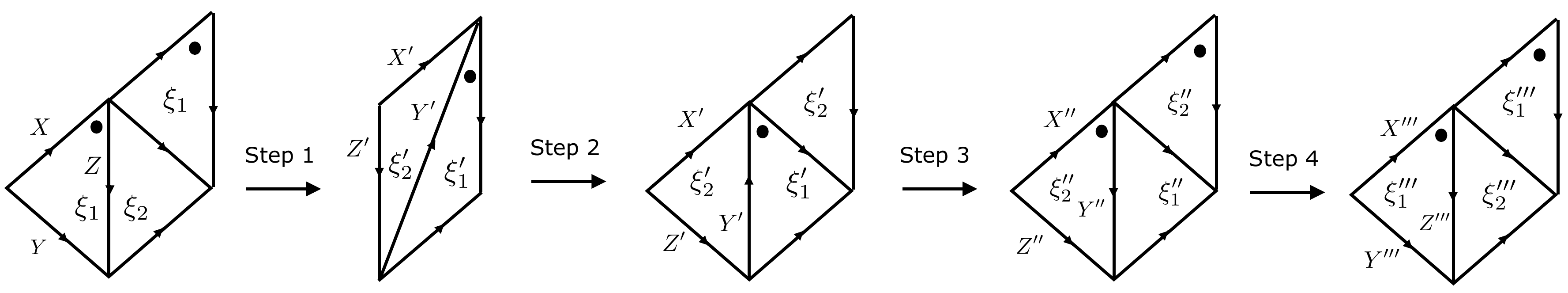}
	\caption{The action of $R_{(1,1)}$ on a dotted triangulation of type $(1,1)$.}
	\label{R1}
\end{figure}

We can work out the expression for $R_{(1,1)}$ in a similar manner (see Fig.~\ref{R1} for 
the four steps\footnote{Compared with the case of $L_{(1,1)}$, one needs to have an extra exchange of $\xi_1$ and $\xi_2$ in the 
Step 4.}):
\begin{align}\label{R1_formula}
\begin{split}
R_{(1,1)} \colon \quad \left(\begin{array}{l}
X \\ Y\\  Z\\ \xi_1 \\ \xi_2
\end{array}\right)
\to \left(
\begin{array}{l}
 X (1 + Y^{-1} + \xi_1\xi_2 \sqrt{Y^{-1}})^{-2} \\
Z (1+ Y + \xi_1\xi_2 \sqrt{Y})^{2}  \\
 Y^{-1} \\
 (1+Y)^{-1/2} (\xi_1\sqrt{Y}+\xi_2) \\
 (1+Y)^{-1/2} (-\xi_1+\xi_2\sqrt{Y}) \\
\end{array}
\right) \;.\\ \smallskip \\
 \end{split}
\end{align}

Notice that when we disregard the  odd coordinates, the spin structure dependence drops out
and the resulting coordinate transformations coincide with those in the non-supersymmetric case, studied previously in \cite{Terashima:2011qi,Terashima:2011xe}.

Since we already know the expression for $L$ and $R$ for all the possible spin structures,
we can derive expression for $S$ in \eqref{ST}:
\begin{align}
\begin{split}
S_{(1,1)}\colon \qquad
\left(\begin{array}{c}
X \\ Y\\  Z\\ \xi_1 \\ \xi_2
\end{array}\right)
\to \left(
\begin{array}{l}
 (1+Z^{-1}+\xi_1\xi_2\sqrt{Z^{-1}})^{-2} Y\\
 (1+Z+\xi_1\xi_2\sqrt{Z})^2 X\\
  Z^{-1} \\
 (1+Z)^{-1/2} (-\xi_1+\xi_2\sqrt{Z}) \\
 (1+Z)^{-1/2} (-\xi_1\sqrt{Z}-\xi_2)
\end{array}
\right) \;.
\end{split}
\end{align}

We can now verify the mapping class group relations satisfied by the $S$ and $T$ generators.
We find
\begin{align}
\begin{split}
(S_{(1,1)})^2 \colon \quad
(X, Y, Z, \xi_1, \xi_2)\to (X, Y, Z, -\xi_2, \xi_1) \;, 
\end{split}
\end{align}
and hence $(S_{(1,1)})^4$ is given by
\begin{equation} \label{S4}
(S_{(1,1)})^4 \colon \quad
(X, Y, Z, \xi_1, \xi_2)\to (X, Y, Z, -\xi_1, -\xi_2)\;.
\end{equation}
Similarly, we can verify that 
\begin{equation} \label{ST3}
(S_{(1,1)} T_{(1,1)})^3 \colon \quad
(X, Y, Z, \xi_1, \xi_2)\to (X, Y, Z, -\xi_2, \xi_1) \;, 
\end{equation}
and
\begin{equation} \label{ST6}
(S_{(1,1)} T_{(1,1)})^6 \colon \quad
(X, Y, Z, \xi_1, \xi_2)\to (X, Y, Z, -\xi_1, -\xi_2)\;.
\end{equation}

In the group $\SL(2, \mathbb{Z})$ one has the relation $S^4=(ST)^6=1$,
however \eqref{S4} and \eqref{ST6} shows that these elements are 
represented non-trivially by an operation of order $2$ acting only on odd variables. 
This suggests that the actual mapping class group relevant for our problem
is the double cover of $\SL(2,\mathbb{Z})$, namely the 
metaplectic group $\Mp(2,\mathbb{Z})$.

The appearance of the metaplectic group can be understood as follows (cf.\ \cite{Pantev:2016nze}).
Let us choose a flat complex coordinate of the 2-dimensional torus $z$, 
with the identification $z\sim z+ m +n\tau$ for integers $m,n$ and the torus modulus $\tau$.
Now, the fermions  takes values in the spinor bundle, and hence transforms as the square root of the one-form
$\sqrt{dz}$: $\sqrt{dz}\to \pm \sqrt{dz} /\sqrt{cz+d}$, with an extra sign ambiguity. This sign gives 
precisely the definition of the metaplectic group,
which is a double cover of the $\SL(2,\mathbb{Z})$ group.

\subsection{Mapping Class Group: Even Spin Structure}\label{sec.even}

Let us next discuss even spin structures. Instead of repeating the manipulations as in Figs.~\ref{L1} and \ref{R1},
we proceed as follows.

Recall that the choice of the spin structure is needed to resolve the sign ambiguity in the choice of the odd variables $\xi_1, \xi_2$,
and hence these are the only ambiguities involved when changing the spin structure. We can represent these ambiguities by sign flips
\begin{align}\label{rho_123}
\begin{aligned}
&\rho_1: ( \xi_1, \xi_2)\to  ( -\xi_1, \xi_2)\;,\\
&\rho_2: ( \xi_1, \xi_2)\to  ( \xi_1, -\xi_2) \;,\\
&\rho_{12}: ( \xi_1, \xi_2)\to  (-\xi_1, -\xi_2) \;,
\end{aligned}
\end{align}
with all the even variables $X, Y, Z$ unchanged.
We can use one of these operators before and/or after to the maps $L_{(1,1)}$ and $R_{(1,1)}$ derived previously.
Moreover, these signs should still be consistent with the mapping class group relations.

It is not difficult to identify the sign rules
which automatically satisfy the mapping class group relations:
\begin{align}\label{rel_L}
\begin{aligned}
& L_{(1,0)} = \rho_{12}^{-1} L_{(1,1)} \rho_2\;, && L_{(0,1)} = \rho_1^{-1} L_{(1,1)} \rho_1\;, &&L_{(0,0)} = \rho_2^{-1} L_{(1,1)} \rho_{12}\;,  \\
 & R_{(1,0)} = \rho_2^{-1} R_{(1,1)} \rho_2\;,&& R_{(0,1)} = \rho_{12}^{-1} R_{(1,1)} \rho_1\;, &&R_{(0,0)} = \rho_1^{-1} R_{(1,1)} \rho_{12} \;,
\end{aligned}
\end{align}

\begin{figure}[htbp]
	\centering
	\scalebox{0.65}{
		\begin{tikzpicture}[every node/.style={}]
		\node (a11) at (3,3)  {$(1,1)$};
		\node (a01) at (0,3)  {$(0,1)$};
		\node (a10) at (3,0)  {$(1,0)$};
		\node (a00) at (0,0)  {$(0,0)$};
		\path[<-] 
		(a11) edge [in=90, out=150, loop, above left] node {$L_{(1,1)}$} ()
		(a11) edge [in=300, out=0, loop, above right] node {$R_{(1,1)}$} ()
		(a01) edge [in=130, out=50, loop, above] node {$L_{(0,1)}$} ()
		(a10) edge [in=330, out=30, loop, right] node {$R_{(1,0)}$} ()
		(a00) edge [bend left] node [below] {$L_{(1,0)}$} (a10) 
		(a10) edge [bend left] node [below] {$L_{(0,0)}$} (a00) 
		(a00) edge [bend right] node [right] {$R_{(0,1)}$} (a01) 
		(a01) edge [bend right] node [left] {$R_{(0,0)}$} (a00) 
		;
		\node (a11) at (13,3)  {$(1,1)$};
		\node (a01) at (10,3)  {$(0,1)$};
		\node (a10) at (13,0)  {$(1,0)$};
		\node (a00) at (10,0)  {$(0,0)$};
		\draw (6.5,-1)--(6.5,5);
		\path[<-] 
		(a11) edge [in=90, out=150, loop, above left] node {$T_{(1,1)}$} ()
		(a11) edge [in=300, out=0, loop, above right] node {$S_{(1,1)}$} ()
		(a01) edge [in=130, out=50, loop, above] node {$T_{(0,1)}$} ()
		(a00) edge [in=150, out=210, loop, left] node {$S_{(0,0)}$} ()
		(a00) edge [bend left] node [below] {$T_{(1,0)}$} (a10) 
		(a10) edge [bend left] node [below] {$T_{(0,0)}$} (a00) 
		(a01) edge [] node [left] {$S_{(1,0)}$} (a10) 
		(a10) edge [bend right] node [right] {$S_{(0,1)}$} (a01) 
		;
		\end{tikzpicture}
	}
	\caption{Orbits of spin structures under the mapping class group transformations $L, R$ and $S, T$. Here the notation $L_{(1,0)}$ represents the $L$ action 
		on the super \Teichmuller space with $(1,0)$ spin structure.}
	\label{map_new}
\end{figure}
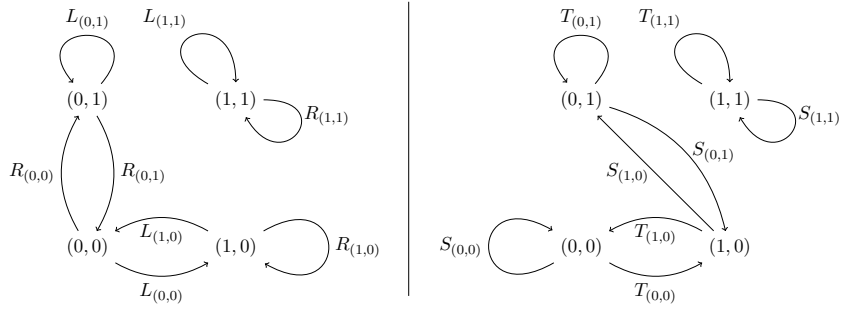

More explicitly,
\begin{align}\label{L234_formula}
\begin{split}
L_{(1,0)}\colon \qquad 
\left(\begin{array}{l}
X \\ Y\\  Z\\ \xi_1 \\ \xi_2
\end{array}\right)
\to \left(
\begin{array}{l}
  Z (1+ X^{-1} - \xi_1\xi_2 \sqrt{X^{-1}})^{-2}\\
 Y (1 + X - \xi_1\xi_2 \sqrt{X})^2 \\
 X^{-1} \\
 (1+X)^{-1/2} (-\xi_1-\xi_2\sqrt{X}) \\
 (1+X)^{-1/2} (-\xi_1\sqrt{X}+\xi_2) 
\end{array}
\right) \;,\\ \smallskip\\
L_{(0,1)}\colon \qquad 
\left(\begin{array}{l}
X \\ Y\\  Z\\ \xi_1 \\ \xi_2
\end{array}\right)
\to \left(
\begin{array}{l}
 Z (1+ X^{-1} - \xi_1\xi_2 \sqrt{X^{-1}})^{-2}\\
Y (1 + X - \xi_1\xi_2 \sqrt{X})^2 \\
X^{-1}\\
 (1+X)^{-1/2} (\xi_1+\xi_2\sqrt{X}) \\
 (1+X)^{-1/2} (-\xi_1\sqrt{X}+\xi_2)  
\end{array}
\right)\;, \\ \smallskip\\
L_{(0,0)}\colon \qquad 
\left(\begin{array}{l}
X \\ Y\\  Z\\ \xi_1 \\ \xi_2
\end{array}\right)
\to \left(
\begin{array}{l}
Z (1+ X^{-1} +\xi_1\xi_2 \sqrt{X^{-1}})^{-2} \\
 Y (1 + X + \xi_1\xi_2 \sqrt{X})^2  \\
X^{-1}\\
 (1+X)^{-1/2} (-\xi_1+\xi_2\sqrt{X})\\
 (1+X)^{-1/2} (\xi_1\sqrt{X}+\xi_2) 
\end{array}
\right) \;,
 \end{split}
\end{align}
and
\vspace{-20pt}
\begin{align}\label{R234_formula}
\begin{split}
R_{(1,0)}\colon \quad
\left(\begin{array}{l}
X \\ Y\\  Z\\ \xi_1 \\ \xi_2
\end{array}\right)
\to \left(
\begin{array}{l}
X (1+ Y^{-1} - \xi_1\xi_2 \sqrt{Y^{-1}})^{-2} \\
Z (1+ Y - \xi_1\xi_2 \sqrt{Y})^{-2} \\
 Y^{-1} \\
 (1+Y)^{-1/2} (+\xi_1\sqrt{Y}-\xi_2) \\
(1+Y)^{-1/2}  (+\xi_1+\xi_2\sqrt{Y}) 
\end{array}
\right) \;,
\\  
\smallskip
\\
R_{(0,1)}\colon \quad
\left(\begin{array}{l}
X \\ Y\\  Z\\ \xi_1 \\ \xi_2
\end{array}\right)
\to \left(
\begin{array}{l}
X (1+ Y^{-1} - \xi_1\xi_2 \sqrt{Y^{-1}})^{-2} \\
Z (1+ Y - \xi_1\xi_2 \sqrt{Y})^{-2} \\
 Y^{-1} \\
 (1+Y)^{-1/2} (\xi_1\sqrt{Y}-\xi_2) \\
(1+Y)^{-1/2}  (-\xi_1-\xi_2\sqrt{Y}) 
\end{array}
\right)\;, 
\\ \smallskip \\
R_{(0,0)}\colon \quad
\left(\begin{array}{l}
X \\ Y\\  Z\\ \xi_1 \\ \xi_2
\end{array}\right)
\to \left(
\begin{array}{l}
X (1+ Y^{-1} + \xi_1\xi_2 \sqrt{Y^{-1}})^{-2} \\
Z (1+ Y + \xi_1\xi_2 \sqrt{Y})^{-2} \\
 Y^{-1} \\
 (1+Y)^{-1/2} (\xi_1\sqrt{Y}+\xi_2) \\
(1+Y)^{-1/2}  (\xi_1-\xi_2\sqrt{Y}) 
\end{array}
\right) \;.
 \end{split}
\end{align}
Note that in each of these cases the product $\xi_1\xi_2$ is preserved under the transformation up to a sign.

In these expressions the choice of $\rho_1, \rho_2, \rho_{12}$ is correlated with the choice of the spin structure:
\begin{equation}\label{phipsi}
\begin{aligned}
&\rho_1: (0,1) \rightarrow (1,1) \;,\\
&\rho_2: (1,0) \rightarrow (1,1) \;,\\
&\rho_{12}: (0,0) \rightarrow (1,1) \;.
\end{aligned}
\end{equation}

For example suppose that we wish to obtain $L_{(0,0)}$, which maps the $(0,0)$ spin structure to the $(1,0)$ spin structure as through the relation $L_{(0,0)} = \rho_2^{-1} L_{(1,1)} \rho_{12}$. We can explain this relation as follows:
One first maps the $(0,0)$ spin structure by $\rho_{12}$ to obtain the $(1,1)$ spin structure,
so that one can apply $L_{(1,1)}$. 
Recall that $L_{(1,1)}$ maps the $(1,1)$ spin structure back to itself. Then we need to transform the 
resulting $(1,1)$ spin structure to the $(1,0)$ spin structure by ${\rho_2}^{-1}$. We can use the similar strategy for the rest of the relations. The relations are summarized on the left side of  Fig.~\ref{map_new}.

The rule \eqref{phipsi} can be regarded as a representation of $\mathbb{Z}_2\oplus \mathbb{Z}_2 \simeq H^1(\Sigma, \mathbb{Z}_2)$,
and this ensures the consistency with the mapping class group relation.
This can also be checked explicitly, by using the relations 
\begin{align}
\begin{split}
&S_{(1,0)} = (L^{-1})_{(0,1)} R_{(0,0)} (L^{-1})_{(1,0)}=(L_{(0,1)})^{-1} R_{(0,0)} (L_{(0,0)})^{-1} \;, \\
&S_{(0,1)} = (L^{-1})_{(0,0)} R_{(0,1)} (L^{-1})_{(0,1)} =(L_{(1,0)})^{-1} R_{(0,1)} (L_{(0,1)})^{-1} \;, \\
&S_{(0,0)} = (L^{-1})_{(1,0)} R_{(1,0)} (L^{-1})_{(0,0)} =(L_{(0,0)})^{-1} R_{(1,0)} (L_{(1,0)})^{-1} \;.
\end{split}
\end{align}
This gives (recall $T=L$, which already appeared in \eqref{rel_L})
\begin{align}\label{Srelations}
\begin{split}
&S_{(1,0)} = \rho_1  S_{(1,1)} \rho_2 \;,  \quad S_{(0,1)} = \rho_2  S_{(1,1)} \rho_1  \;, \quad S_{(0,0)} =  S_{(1,1)}  \;, \\
&T_{(1,0)} = T_{(1,1)} \rho_1\;,  \quad T_{(0,1)} = \rho_1 T_{(1,1)} \rho_1\;, \quad T_{(0,0)} = \rho_2 T_{(1,1)} \rho_{12}\;.
\end{split}
\end{align}
For example, if we want to evaluate $S^4$ starting with the type $(1,0)$ spin structure,
one needs to evaluate $(S^4)_{(1,0)}=S_{(0,1)} S_{(1,0)} S_{(0,1)} S_{(1,0)}$,
which coincides with $(S^4)_{(1,1)}$. We can similarly compute $(ST)^6$ for all the even structures, 
and find that they all coincide with that of the odd spin structure.
This verifies the expected relations in the metaplectic group $\Mp(2,\mathbb{Z})$.
%
The right side of  Fig.~\ref{map_new} summarizes how $S$ and $T$ map one spin structure to another.

\section{Quantum Super \Teichmuller Theory}\label{sec:quantum}

In this section, we describe the quantization of the super \Teichmuller theory (as discussed in \cite{Aghaei:2015bqi}).
We again discuss the case of the once-punctured torus. First,  in Sec. \ref{5.1} we define the Hilbert space of the theory on the torus and how the quantized super Fock coordinates are represented on it.  We then discuss how the mapping class group generators are represented by suitable operators acting on the Hilbert space in  Sec. \ref{5.2}.

\subsection{Quantization of the Super Fock Space}
\label{5.1}

The Hilbert space associated to a dotted ideal triangulation of a super Riemann surface
is defined as follows. To each dotted triangle $\Delta_v$ 
we associate a Hilbert space $\mathcal{H}(\Delta_v) \equiv \mathcal{H}_v \simeq L^2(\mathbb{R})\otimes\mathbb{C}^{1|1}$. Therefore, the Hilbert space $\mathcal{H}_\sigma(\Sigma_{1,1}) \equiv \mathcal{H}_\sigma$ associated to the torus with one of the four spin structures $\sigma=(0,0),(0,1),(1,0),(1,1)$ is the tensor product of the spaces for each triangle. For once-punctured torus in particular, we have
\begin{equation}
\mathcal{H}_\sigma = \mathcal{H}_1 \otimes \mathcal{H}_2 \;.
\end{equation}

The super Fock coordinates $x = \log(X), y = \log(Y), z=\log(Z)$, which are logarithms of the super Fock coordinates $X,Y,Z$ introduced in \eqref{conformalinvariantt}, get quantized to self-adjoint operators on the Hilbert spaces $\mathcal{H}_v$. 
The even coordinates $\mathsf{x}, \mathsf{y}, \mathsf{z}$ are replaced by operators satisfying canonical commutation relations 
\begin{equation}
\left[\mathsf{x},\mathsf{y}\right]= \left[\mathsf{y},\mathsf{z}\right]= \left[\mathsf{z},\mathsf{x}\right]= -8\pi i \ub^2 \;.
\end{equation}
The algebra of those even coordinates admits a central element
\begin{equation}
\mathsf{h} \coloneqq  \mathsf{x}+\mathsf{y}+\mathsf{z}\;.
\end{equation}
One can hence consider a decomposition of $\mathcal{H}$ into spaces on which $\mathsf{h}$ is diagonal 
\begin{equation}
\mathcal{H}_\sigma = \bigoplus_{h\in \mathbb{R}
} \mathcal{P}_h \;,
\end{equation}
where $\mathcal{P}_h \simeq L^2(\mathbb{R})\otimes (\mathbb{C}^{1|1})^{\otimes2}$. On $\mathcal{P}_h$ two of the observables $\mathsf{x}, \mathsf{y}, \mathsf{z}$ (we will choose $\mathsf{x}$ and $\mathsf{y}$) are represented on $L^2(\mathbb{R})$ as multiplication and differentiation operators. In the classical limit $\ub\to0$, the operators $\mathsf{x}, \mathsf{y}, \mathsf{z}$ give their classical counterparts $x,y,z$ as one would expect. The odd coordinates $\xi_i$ become operators acting on $\mathcal{H}_{\sigma}$ of the form
\begin{align}
\pi_h(\xi_1) &= i\sqrt{ q^\frac{1}{2}-q^{-\frac{1}{2}} } (\kappa \otimes \mathbb{I}_2)\;, && \pi_h(\xi_2) = i\sqrt{ q^\frac{1}{2}-q^{-\frac{1}{2}} } (\mathbb{I}_2 \otimes \kappa)\;,
\end{align}
where $\kappa$ is a $(1|1)\times(1|1)$ matrix acting on $\mathbb{C}^{1|1}$
\begin{equation}
\kappa = \bigg( \begin{array}{cc} 0 & 1 \\ 1 & 0 \end{array} \bigg)\;,
\end{equation}
$\mathbb{I}_2$ is the $(1|1)$-dimensional identity matrix, and $q\coloneqq e^{i\pi \ub^2}$. One finds that $\xi_i$ satisfy anti-commutation relations 
\begin{align}
\{\xi_i, \xi_i\}= - 2 \left( q^\frac{1}{2}-q^{-\frac{1}{2}} \right) \;, && \{\xi_1,\xi_2\} =0 \;,
\end{align}
and commute with all even operators 
\begin{align}
& [\mathsf{x}, \xi_i]=[\mathsf{y},\xi_i] = [\mathsf{z},\xi_i] = 0 \;.
\end{align}
Summarizing, the quantized super Fock variables are represented on the space 
\begin{align}
\mathcal{P}_h = \text{span}\left\{ |x\rangle \otimes \left( \begin{array}{c} a \\ b \end{array} \right) \otimes \left( \begin{array}{c} c \\ d \end{array} \right) \right\}_{x\in\mathbb{R}, a,b,c,d\in\mathbb{C}},
\end{align}
as follows:
\begin{align}
\begin{split}
\pi_h(\mathsf{x}) &= x \otimes \mathbb{I}_2 \otimes \mathbb{I}_2 \;, \\
\pi_h(\mathsf{y}) &= 8 \pi i \ub^2 \frac{\ud}{\ud x} \otimes \mathbb{I}_2 \otimes \mathbb{I}_2 \;, \\
\pi_h(\mathsf{z}) &= \left(h - x - 8 \pi i \ub^2 \frac{\ud}{\ud x} \right) \otimes \mathbb{I}_2 \otimes \mathbb{I}_2 \;, \\
\pi_h(\xi_1) &= i\sqrt{q^\frac{1}{2}-q^{-\frac{1}{2}}} \otimes \kappa \otimes \mathbb{I}_2\;, \\
\pi_h(\xi_2) &= i\sqrt{q^\frac{1}{2}-q^{-\frac{1}{2}}} \otimes \mathbb{I}_2 \otimes \kappa \;.
\end{split}
\end{align}

\subsection{Mapping Class Group Generators}
\label{5.2}

Now, we will realize the Dehn twists $L, R$ described in Sec.~\ref{sec.MCG} as linear operators $\mathsf{L}, \mathsf{R}$ acting on the Hilbert space. 
A coordinate transformation maps one spin structure $\sigma$ to another $\sigma'$,
and we promote it to an operator $\mathsf{U}_{\sigma'\sigma}: \mathcal{H}_\sigma \to \mathcal{H}_{\sigma'}$ between the corresponding Hilbert spaces.
Since we already know the classical coordinate transformation, we know a transformation rule of the form ${w}'{}^{\jmath}= W_{\sigma'\sigma}^{\jmath}(\{w^{\imath}\})$,
where $w^{\imath}$ and $w^{\jmath}$ are coordinates corresponding for the Hilbert spaces  $\mathcal{H}_\sigma$ and $\mathcal{H}_{\sigma'}$.
The unitary operator  $\mathsf{U}_{\sigma'\sigma}^{}$ representing these changes of coordinates on the quantum level satisfy
\begin{equation}\label{transfgeneral}
\mathsf{U}_{\sigma'\sigma}^{-1} \cdot {\mathsf{w}'}^{\jmath} \cdot \mathsf{U}_{\sigma'\sigma}^{} = 
\mathsf{W}_{\sigma'\sigma}^{\jmath}(\{\mathsf{w}_{\imath}\})\,,
\end{equation}
and should be consistent with the classical transformation rule in the classical limit.
This requirement is expected to characterize the operators $ \mathsf{U}_{\sigma'\sigma}$ uniquely up to a multiplicative coefficient. 

We will start by considering the quantization of the coordinate transformations \eqref{L1_formula} and \eqref{R1_formula} given by the quantized Dehn twists associated to the $(1,1)$ spin structure\\ $\mathsf{L}_{(1,1)}: \mathcal{H}_{(1,1)} \to \mathcal{H}_{(1,1)}$:
\begin{align}\label{quantum_L_transformation_1}
\begin{split}
\mathsf{L}_{(1,1)}^{-1}\, e^\mathsf{x} \, \mathsf{L}_{(1,1)} &= (1+q^2 e^{-\mathsf{x}}+q e^{-\mathsf{x}/2} \xi_1\xi_2)^{-1}(1+q^6 e^{-\mathsf{x}}+q^3 e^{-\mathsf{x}/2}\xi_1\xi_2)^{-1} e^{\mathsf{z}}\;, \\
\mathsf{L}_{(1,1)}^{-1}\,  e^\mathsf{y}\,  \mathsf{L}_{(1,1)} &= (1+q^2 e^{\mathsf{x}}+q e^{\mathsf{x}/2} \xi_1\xi_2) (1+q^6 e^\mathsf{x} +q^3 e^{\mathsf{x}/2} \xi_1\xi_2) e^{\mathsf{y}}\;, \\
\mathsf{L}_{(1,1)}^{-1} \, e^\mathsf{z}\,  \mathsf{L}_{(1,1)} &= e^{-\mathsf{x}} \;, \\
\mathsf{L}_{(1,1)}^{-1}\,  \xi_1 e^{\mathsf{y}/4}\,  \mathsf{L}_{(1,1)} &= (\xi_1 - q^{1/2} e^{\mathsf{x}/2} \xi_2) e^{\mathsf{y}/4} \;, \\
\mathsf{L}_{(1,1)}^{-1}\,  \xi_2 e^{\mathsf{y}/4} \, \mathsf{L}_{(1,1)} &= (\xi_2 + q^{1/2} e^{\mathsf{x}/2} \xi_1) e^{\mathsf{y}/4} \;,
\end{split}
\end{align}
and $\mathsf{R}_{(1,1)}: \mathcal{H}_{(1,1)} \to \mathcal{H}_{(1,1)}$:
\begin{align}
\begin{split}
\mathsf{R}_{(1,1)}^{-1} \, e^\mathsf{x} \, \mathsf{R}_{(1,1)} &= e^\mathsf{x} (1 + q^{-6} e^\mathsf{-y} + q^{-3} \xi_1\xi_2 e^\mathsf{-y/2})^{-1} (1 + q^{-2} e^\mathsf{-y} + q^{-1} \xi_1\xi_2 e^\mathsf{-y/2})^{-1}\;, \\
\mathsf{R}_{(1,1)}^{-1} \, e^\mathsf{y} \, \mathsf{R}_{(1,1)} &= e^\mathsf{z} (1+ q^{-6}e^\mathsf{y} + q^{-3}\xi_1\xi_2 e^\mathsf{y/2})(1+ q^{-2}e^\mathsf{y} + q^{-1}\xi_1\xi_2 e^\mathsf{y/2}) \;, \\
\mathsf{R}_{(1,1)}^{-1} \, e^\mathsf{z} \, \mathsf{R}_{(1,1)} &= e^\mathsf{-y}\;, \\
\mathsf{R}_{(1,1)}^{-1}\, \xi_1 e^{\mathsf{y}/4} \, \mathsf{R}_{(1,1)} &= (\xi_2+ q^{1/2} \xi_1 e^{\mathsf{y}/2} ) e^{\mathsf{z}/4}\;, \\
\mathsf{R}_{(1,1)}^{-1}\, \xi_2 e^{\mathsf{y}/4}\, \mathsf{R}_{(1,1)} &= (-\xi_1+q^{1/2} \xi_2e^{\mathsf{y}/2}) e^{\mathsf{z}/4} \;.\label{quantum_R_transformation_5}
\end{split}
\end{align}
The operators $\mathsf{L}_{(1,1)}, \mathsf{R}_{(1,1)}$ implementing the above transformations can be constructed as follows
\begin{align}
\begin{split}
\mathsf{L}_{(1,1)} &= e^{\frac{1}{16\pi i \ub^2}(\mathsf{x}+\mathsf{z})^2} \textbf{e}^{-1}\left(\frac{\mathsf{x}}{2\pi \ub}\right) \;, \\
\mathsf{R}_{(1,1)} &= e^{-\frac{1}{16\pi i \ub^2}(\mathsf{y}+\mathsf{z})^2} \textbf{e}\left(-\frac{\mathsf{y}}{2\pi \ub}\right) \;.
\end{split}
\end{align}
Here $\textbf{e}$ is a function-valued matrix
\begin{align} \label{def_e}
\begin{split}
\textbf{e}(u) & \coloneqq \frac{1}{2} \left[ e_\R(u)  (\mathbb{I}_2\otimes \mathbb{I}_2 -i \kappa\otimes \kappa)
+e_\NS(u)(\mathbb{I}_2\otimes \mathbb{I}_2 + i \kappa\otimes \kappa) \right] \;,\\
\textbf{e}^{-1}(u) &= \frac{1}{2} \left[ e_\R(u)^{-1}  (\mathbb{I}_2\otimes \mathbb{I}_2 -i \kappa\otimes \kappa)
+e_\NS(u)^{-1}(\mathbb{I}_2\otimes \mathbb{I}_2 + i \kappa\otimes \kappa) \right]  \;,
\end{split}
\end{align}
where the special functions $e_\R,e_\NS$ are the supersymmetric analogs of the Faddeev's quantum dilogarithm (cf.\ \cite{Fukuda:2002bv,Hadasz:2007wi})
\begin{align}\label{e_SUSY}
\begin{split}
e_\R(x) & \coloneqq e_\ub\left(\frac{x+i(\ub-\ub^{-1})\slash2}{2}\right)e_\ub\left(\frac{x-i(\ub-\ub^{-1})\slash2}{2}\right),\\
e_\NS(x) & \coloneqq e_\ub\left(\frac{x+i(\ub+\ub^{-1})\slash2}{2}\right)e_\ub\left(\frac{x-i(\ub+\ub^{-1})\slash2}{2}\right) ,
\end{split}
\end{align}
and the quantum dilogarithm function $e_\ub(x)$ \cite{Faddeev:1993pe,Faddeev:1993rs,Faddeev:1995nb} is defined by the following integral representation
\begin{equation}
e_\ub(x) \coloneqq \exp \left[\int_{\mathbb{R}+i0} \frac{\ud w}{w} \frac{e^{-2ixw}}{4 \sinh(w\ub)\sinh(w\slash \ub)}\right].
\end{equation}
In particular, the coordinate transformations \eqref{quantum_L_transformation_1} are satisfied thanks to the shift property of 
the function $\textbf{e}$
\begin{align}
\begin{split}
&\textbf{e}\left(x-\frac{i\ub}{2}\right)(\kappa\otimes \mathbb{I}_2) = (\kappa\otimes\mathbb{I}_2 - e^{\pi \ub x} \mathbb{I}_2\otimes\kappa) \,\textbf{e}\left(x+\frac{i\ub}{2}\right) \;, \\
&\textbf{e}\left(x-\frac{i\ub}{2}\right)(\mathbb{I}_2\otimes\kappa) = (\mathbb{I}_2\otimes\kappa + e^{\pi \ub x} \kappa\otimes\mathbb{I}_2) \,\textbf{e}\left(x+\frac{i\ub}{2}\right) \;.
\end{split}
\end{align}
These equations follow from the shift properties
\begin{align}
\begin{split}
e_\R\left(x-\frac{i \ub^{\pm1}}{2}\right) &= (1+ie^{\pi \ub^{\pm1}x}) \, e_\NS\left(x+\frac{i\ub^{\pm1}}{2}\right) \;, \\
e_\NS\left(x-\frac{i \ub^{\pm1}}{2}\right) &= (1-ie^{\pi \ub^{\pm1}x})\, e_\R\left(x+\frac{i\ub^{\pm1}}{2}\right) \;,
\end{split}
\end{align}
which is implied by a similar relation for the non-supersymmetric quantum dilogarithm:
\begin{align}
e_\ub \left(x-\frac{i \ub^{\pm1}}{2}\right) &= (1+e^{2\pi \ub^{\pm1}x}) \, e_\ub\left(x+\frac{i\ub^{\pm1}}{2}\right) \;.
\end{align}

\subsection{Change of Kasteleyn Orientations}

We now describe operators changing the Kasteleyn orientations, as well as those changing spin structures.

For a given spin structure, any two Kasteleyn orientations are related by push outs.
Recall that the push out $\beta_{12}$ of Fig.~\ref{pushleftdot}
flips the sign of the one of the odd variables $\xi_2$,
while preserving the remaining odd variable $\xi_1$ as well as all the 
even variables $x, y, z$. The quantum version of this operator can be identified to be
\begin{align}\label{BM}
\mathsf{B}_{12}=  \mathbb{I}_2 \otimes M \;,
\quad 
M\coloneqq \left(\begin{array}{cc} 1 & 0 \\ 0 & -1\end{array} \right) \;.
\end{align}
We can easily verify the expected property
\begin{align}
\begin{split}
&\mathsf{B}_{12}^{-1} X \mathsf{B}_{12}=X \;, \quad
\mathsf{B}_{12}^{-1} Y \mathsf{B}_{12}=Y \;, \quad
\mathsf{B}_{12}^{-1} Z \mathsf{B}_{12}=Z \;, \\
&\mathsf{B}_{12}^{-1} \xi_1 \mathsf{B}_{12}=\xi_1 \;, \quad
\mathsf{B}_{12}^{-1} \xi_2 \mathsf{B}_{12}=-\xi_2 \;.
\end{split}
\end{align}

We can also discuss changes of spin structures.
In order to describe the operators $\mathsf{L}, \mathsf{R}$ for even spin structures, we introduce ``spin structure changing operators''
(whose classical analogs were introduced in \eqref{phipsi}):
\begin{align}
\srho_{12}: \mathcal{H}_{(0,0)} \rightarrow \mathcal{H}_{(1,1)} \;, &&  \srho_1: \mathcal{H}_{(0,1)} \rightarrow \mathcal{H}_{(1,1)} \;, && \srho_2: \mathcal{H}_{(1,0)} \rightarrow \mathcal{H}_{(1,1)} \;,
\end{align}
given by
\begin{align} \label{M_q}
\srho_{12} = M \otimes M \;, && \srho_1 = M\otimes \mathbb{I}_2 \;, && \srho_2 =\mathbb{I}_2\otimes M\;,
\end{align}
where $M$ was defined previously in \eqref{BM}. Then, the remaining quantized Dehn twists are related to the one described above as follows 
\begin{align}\label{L_rel_q}
\begin{aligned}
 &\mathsf{L}_{(1,0)} = \srho_{12}^{-1} \mathsf{L}_{(1,1)} \srho_2\;,  \quad  &&\mathsf{L}_{(0,1)} = \srho_1^{-1} \mathsf{L}_{(1,1)} \srho_1 \;,  \quad&&\mathsf{L}_{(0,0)} = \srho_2^{-1} \mathsf{L}_{(1,1)} \srho_{12} \;,\\
&\mathsf{R}_{(1,0)} = \srho_2^{-1} \mathsf{R}_{(1,1)} \srho_2\;,
 \quad &&\mathsf{R}_{(0,1)} = \srho_{12}^{-1} \mathsf{R}_{(1,1)} \srho_1 \;,  \quad &&\mathsf{R}_{(0,0)} = \srho_1^{-1} \mathsf{R}_{(1,1)} \srho_{12} \;.\end{aligned}
\end{align} 

\section{Partition Functions for Super \Teichmuller Spin TQFT}\label{sec:Z}

\subsection{Definition of the Partition Function}\label{subsec:def_Z}

As already explained in Section \ref{sec:outline}, our basic idea is that the 
trace $\Tr(\hvarphi)$ of the mapping class group action $\hvarphi$ inside the Hilbert space
should basically be the partition function of the 
super \Teichmuller spin TQFT. 

The spin structure we have used so far is a spin structure of the 2-manifold.
For the super \Teichmuller spin TQFT, however, it should be that we need a spin structure of the 3-manifold (mapping torus). To discuss this, let us note that the fundamental group of the mapping torus \eqref{mt_def} is given by
\begin{align}
\pi_1\left(M \right)= \pi_1 (\Sigma)  \rtimes_{\varphi_{*}} \mathbb{Z}\;,
\end{align}
where $\varphi_{*}: \pi_1(\Sigma)\to \pi_1(\Sigma)$ is the map induced from $\varphi\in \mathrm{Aut}(\Sigma)$.
In other words, $\pi_1\left(M \right)$ is given by $\pi_1 (\Sigma)$, with an extra generator $\gamma$ added 
and with extra relations
\begin{align}
\pi_1(M)= \left\{\gamma^{-1} \alpha \gamma= \varphi_{*}(\alpha)\;, \quad \alpha\in \pi_1(\Sigma)  \right\}\;.
\label{gag}
\end{align}
Note that the space of spin structures is an affine space over $H^1(M, \mathbb{Z}_2)$,
which is a $\mathbb{Z}_2$-reduction of $H^1(M, \mathbb{Z})$,
which in turn is the abelianization of $\pi_1(M, \mathbb{Z})$. This makes it clear that, in order to discuss spin structures on mapping tori
we need to \\
(1) choose the spin structure of $\Sigma$ which is kept fixed under $\varphi$
and\\
 (2) take into account an extra $\mathbb{Z}_2$-choice of the spin structure,
corresponding to the extra cycle $\gamma$ (i.e.\ the extra circle $S^1$, the base of the mapping torus).

Let us first discuss point (2).
We claim that we can distinguish the two spin structures along the $S^1$-direction
by introducing projection operators $P_\pm$:
\begin{equation}
P_\pm \coloneqq \frac{1}{2} \left[\mathbb{I}_2\otimes \mathbb{I}_2 \mp i \kappa\otimes\kappa\right].
\end{equation}
They satisfy the canonical relations for projection operators
\begin{align}
P_\pm^2 = \mathbb{I}_2 \;, \quad P_+ P_- = P_- P_+ = 0 \;, \quad P_{+} + P_{-} =\mathbb{I}_2\otimes \mathbb{I}_2 \;.
\end{align}
We then consider two partition functions by the 
trace with the projection operators inserted:
\begin{align}
\begin{split}
\Tr_{\rm NS}(\hvarphi) & \coloneqq\Tr (P_{-} \hvarphi) \;, \\
\Tr_{\rm R}(\hvarphi) &\coloneqq \Tr (P_{+} \hvarphi) \;.
\end{split}
\end{align}

We claim that the two choices represents the two choices of the 
spin structures. Indeed, since we have the relation
\begin{align}
\begin{split}
&P_+ \left[ a (\mathbb{I}_2\otimes \mathbb{I}_2 -i \kappa\otimes \kappa) +b (\mathbb{I}_2\otimes \mathbb{I}_2 +i \kappa\otimes \kappa)  \right] = a P_+\;, \\
&P_- \left[ a (\mathbb{I}_2\otimes \mathbb{I}_2 -i \kappa\otimes \kappa) +b (\mathbb{I}_2\otimes \mathbb{I}_2 +i \kappa\otimes \kappa) \right] = b P_- \;.
\end{split}
\end{align}
We find that when computing $\Tr_{\rm NS}(\hvarphi)$ and $\Tr_{\rm R}(\hvarphi)$
we can replace $\textbf{e}(u)$ and $\textbf{e}^{-1}(u)$ in \eqref{def_e} by
\begin{align}
\begin{aligned}
\textbf{e}(u) &\leadsto  e_\NS(u) P_{-}\;, \quad &&\textbf{e}^{-1}(u) \leadsto  e_\NS(u)^{-1} P_{-}\quad &&(\textrm{for $\Tr_{\rm NS}(\hvarphi)$ }) \;,\\
\textbf{e}(u) &\leadsto  e_\R(u) P_{+}\;, \quad &&\textbf{e}^{-1}(u) \leadsto  e_\R(u)^{-1} P_{+} \quad &&(\textrm{for $\Tr_{\rm R}(\hvarphi)$ })\;.
\end{aligned}
\end{align}
This observation simplifies our computation below considerably.

Let us now come back to point (1) concerning the condition that the spin structure is fixed by $\varphi$.
While in general there can be more than one of such spin structures satisfying this condition,
operators $\hvarphi$ fixing spin structures are related by conjugation by spin structure changing operators \eqref{M_q}.
Since these operators are even, and since these operators either commute with $P_{\pm}$ or exchange $P_+$ and $P_-$, 
their associated partition functions (which are defined by the trace) coincide due to the conjugation invariance of the trace.
This means that we always have two partition functions depending on the choice of R/NS
in the trace.\footnote{One can also try to compute the partition function when the spin structure is not fixed by $\varphi$.
Since this is inconsistent geometrically, we expect that something should go wrong for these cases.
It turns out that the partition functions are zero in these cases.
We can see this in a simple example of $\varphi=L$.
This preserves only two (types $(1,1)$ and $(0,1)$) out of the four spin structures.
The corresponding expressions for $L_{(1,1)}$ and $L_{(0,1)}$ are non-trivial,
and the two expressions are the same,  since according to \eqref{L_rel_q}
$L_{(1,1)}$ and $L_{(0,1)}$ are related by conjugation by $\srho_1$.
By contrast $\Tr( L_{(1,0)})$ and $\Tr(L_{(0,0)})$ vanishes simply because
$\Tr(\srho_1 P_{\pm}),  \Tr(\srho_2 P_{\pm}),  \Tr(\srho_{12} P_{\pm})$
all vanish. 
A similar discussion shows that this vanishing property holds more generally
for an arbitrary element of the mapping class group not preserving the spin structure.}
For this reason it is sufficient below to describe the 
partition function starting with the odd spin structure (of type $(1,1)$),
which is always kept fixed irrespective of the choice of the mapping class group.

For our discussion of the quantum super \Teichmuller theory,
we need to choose both an ideal triangulation and a Kasteleyn orientation on it.
Recall that  different ideal triangulations are related by superflips,
and that different Kasteleyn orientations (for a given spin structure)
are related by push outs.
Since superflips and push outs are represented by unitary operators (which we call $\mathsf{U}$),
such ambiguities have the effect of replacing the operator $\hvarphi$
by $\mathsf{U}^{-1}\hvarphi \mathsf{U}$.
The operators $\mathsf{U}$ were even elements, and also commute with the projection operators $P_{\pm}$.
This means that the two traces, $\Tr(P_{\pm} \hvarphi)$ and $\Tr(P_{\pm} \mathsf{U}^{-1} \hvarphi \mathsf{U})$ coincide,
thanks to the conjugate-invariance of the trace.\footnote{One can instead choose a supertrace which is automatically conjugation invariant.
The resulting expressions $\textrm{Str}(P_{\pm} \hvarphi)$ vanishes, however.}
The partition function as defined by the trace is thus free from the 
ambiguities mentioned above, and depends only on the choice of the spin structure of the 3-manifold, as expected. 

In the following we consider a few concrete examples of the 
mapping class group element $\varphi$ and the associated partition functions.
For $|\Tr(\varphi)|>2$, the mapping class group $\varphi$ is known as pseudo-Anosov 
and the resulting mapping class torus admits a complete hyperbolic structure \cite{MR1402300}.

\subsection{Example: \texorpdfstring{$\varphi=LR$}{varphi=LR}}

Let us consider an example of 
$\varphi=LR=\left(\begin{array}{cc}
2&1 \\
1&1
\end{array}\right)$.
 In this case, the mapping torus $M$
is identified with complement of the figure-eight knot (often denoted by $\mathbf{4}_1$) inside $S^3$: $S^3\setminus \mathbf{4}_1$.
We wish to compute the partition function of this 3-manifold. Since we have two spin structures on this manifold (corresponding to the $\mathbb{Z}_2$ spins around the tubular neighborhood of the knot),
we expect to have two partition functions, and these two will be identified with our partition functions $\Tr_{\rm R/NS}(\mathsf{LR})$.

On the 2-manifold $\Sigma$ the odd spin structure is kept invariant under $\varphi$,
while the three even spin structures are permuted among them. We thus need to choose the odd spin structure
for mapping torus construction.

We can define the trace for the mapping class element $\varphi=LR$ as a trace of the operator~$\mathsf{RL}$
\begin{align}
\Tr_{\textcolor{red}{\rm R/NS}}(\mathsf{LR}) &= \text{Tr}_{ L^2(\mathbb{R})\otimes \mathbb{C}^{2|2}}(P_{\pm} \mathsf{R L}) \nonumber \\
&= \int \ud x\ud y \, e^{-\frac{1}{\mathcolor{red}{8}\hbar}[(h-x)^2-(h-y)^2]} e_{\textcolor{red}{\R/\NS}}\left( -\frac{y}{2\pi b}\right) e_{\textcolor{red}{\R/\NS}}^{-1}\left( \frac{x}{2\pi b}\right) \;,
\label{susy_LR} 
\end{align}
where we denoted 
\begin{align}
\hbar \coloneqq 2 \pi i b^2 \;.
\end{align}
It is worthwhile to compare this result to the calculation for the bosonic case \cite{Terashima:2011xe}:
\begin{align*}
\text{Tr}_{\textrm{non-SUSY}}( \mathsf{LR}) 
&= \int \ud x\ud y \, e^{-\frac{1}{4\hbar}[(h-x)^2-(h-y)^2]} e_\ub\left(-\frac{y}{2\pi b}\right) e_\ub^{-1}\left(\frac{x}{2\pi b}\right) \;.\label{nonsusy_LR} 
\end{align*}
The comparison of the two expressions makes clear that 
the only differences between the non-supersymmetric and supersymmetric cases are (the two are related, as will become clear below):
\begin{screen}
\begin{enumerate}
\item $e_\ub$ is replaced by $e_{\R}$ and $e_{\NS}$
depending on the choice of the spin structure,
\item $\hbar=2\pi i \ub^2$ is replaced by $\hbar/2$.
\end{enumerate}
\end{screen}
In this section we often highlight these  differences in red color.

\bigskip

We can now discuss the asymptotic expansion of the expression around $b=0$ ($\hbar= 0$). For this purpose we need the asymptotic expansion of the 
supersymmetric quantum dilogarithm around $b=0$:\footnote{This can be derived from the Euler-Maclaurin expansion of the quantum dilogarithm function:
\begin{align}
 e_\ub \left(z+m \frac{i \ub}{4} + n \frac{i \ub^{-1}}{4} \right) = \exp\left[ \sum_{k=0}^\infty \frac{B_k(\frac{1}{2}+\frac{m}{4})}{k!} (2i\pi \ub^2)^{k-1} \Li_{2-k}(-i^n e^{2\pi \ub z}) \right] .
\end{align}
}
\begin{align}
\begin{split}
 e_\R(x) &= \exp\left[ \sum_{n=0}^\infty \frac{B_n(\frac{1}{4})}{n!} \hbar^{n-1} (\Li_{2-n}(-i e^{\pi bx})+(-1)^n \Li_{2-n}(i e^{\pi bx})) \right] \;, \\
 e_\NS(x) &= \exp\left[ \sum_{n=0}^\infty (-1)^n \frac{B_n(\frac{1}{4})}{n!} \hbar^{n-1} (\Li_{2-n}(-i e^{\pi bx})+(-1)^n \Li_{2-n}(i e^{\pi bx})) \right] \;,
 \end{split}
\end{align}
where $B_n(x)$ is the $n$-th Bernoulli polynomial and $\Li_{n}(x)$ is the polylogarithm function.

\bigskip

In the following we concentrate on the leading and subleading terms,
which read (using $B_0(x)=1, B_1(x)=x-1/2, \Li_{2}(x)+\Li_{2}(-x)= \Li_{2}(x^2)/2$ and $\Li_{1}(x)=-\ln(1-x)$)
\begin{align}
\begin{split}
 e_{\R/\NS}(x) &= \exp\left[  \frac{1}{\mathcolor{red}{2}\hbar} \Li_{2}(- e^{2\pi \ub x})  \mathcolor{red}{\pm  \frac{1}{4}\ln  \mathcal{L}(x)}+ O(\hbar)\right] \;, 
 \end{split}
\end{align}
where plus/minus sign corresponds to R/NS respectively, and we defined
\begin{align}
\mathcal{L}(x):=  \frac{1+i e^{\pi \ub x}}{1-i e^{\pi \ub x}} \;.
\end{align}
It is again useful to compare this with the expansion of the quantum dilogarithm function itself:
\begin{align}
 e_{\ub}\left(z \right) = \exp\left[  \frac{1}{\hbar} \Li_{2}(-e^{2\pi \ub z})
  + O(\hbar) \right] .
\end{align}
In the leading order, we have again a factor $2$ difference in $\hbar$,
which is consistent with the observation before. What is important is that 
in the subleading order there is a new contribution ($\mathcal{L}(x)$) in the supersymmetric case,
which does not have a counterpart in the non-supersymmetric case.

With the help of these formulas, the partition function can be expanded in the limit $\hbar, b\to 0$ as
\begin{align}
\text{Tr}_{\rm R/NS} (\mathsf{LR}) 
&= \int \ud x\ud y \, e^{\frac{1}{\mathcolor{red}{2} \hbar} V_0(x,y,h) \mathcolor{red}{\pm \frac{1}{4} \ln V_1(x,y)} } (1 + O(\hbar)) \;,
\end{align}
where
\begin{align}
\begin{split}
V_0(x,y,h) & \coloneqq  -\frac{1}{4}(h-x)^2 -  \text{Li}_2(-e^x) + \frac{1}{4} (h-y)^2 +  \text{Li}_2(-e^{-y}) \;, \\
V_1(x,y) & \coloneqq \frac{\mathcal{L}(-y)}{\mathcal{L}(x)} \;. 
\end{split}
 \end{align} 

This integral can be evaluated in the saddle point approximation, namely by extremizing the function $V_0(x,y,h)$.

For the special case of $h=0$, the saddle points are given by
$e^x=e^y=(-1+i\sqrt{3})/2$, and the corresponding critical value $S_0=\Li_2(-e^{2\pi i/3})- \Li_2(-e^{-2\pi i/3})$ of $V_0$ gives the 
complexified volume of the figure eight knot complement. 

We can also keep the general values of $h$. In this case, the saddle point equations 
describe the gluing conditions for two ideal tetrahedra triangulating the figure eight knot complement.
The parameter $h$, identified with the longitude parameter $\mathfrak{l}$ of the boundary torus by a simple relation $\mathfrak{l} \coloneqq  h + i\pi$ \cite{Terashima:2011xe}, describes the one-parameter family of the deformation of the hyperbolic structure \cite{NeumannZagier}. For comparison with literature,
it is useful to fix the meridian parameter $\mathfrak{m}$, which we can achieve by 
a Fourier transformation (cf.\ \cite[Appendix C]{Gang:2015wya})
\begin{align}
\widetilde{\text{Tr}}_{\rm R/NS} (\mathfrak{m})\coloneqq \int d h  \,\, \text{Tr}_{\rm R/NS} (h) \, e^{-\mathfrak{m}\,h/(2 \hbar)} \;.
\end{align}
In this expression, the longitude parameter $\mathfrak{l}$ plays the role of the Lagrange multiplier,
giving rise to the constraint\footnote{This coincides with the expression for the meridian coming from ideal triangulations of the boundary torus of the knot complement \cite{Terashima:2011xe}.}
\begin{align}
\mathfrak{m} = \frac{\partial V_0(x,y,h)}{\partial h}=(x-y)/2 \;.
\end{align}
We can then eliminate the variable $y$ in terms of $x$ and $\mathfrak{m}$,
and evaluate the asymptotic expansion with respect to the remaining integration variable $x$.

The leading piece $V_0$ now reads
\begin{align}
V_0(x, \mathfrak{m})  = -  \text{Li}_2(-e^x) +\text{Li}_2(-e^{-(x-2\mathfrak{m}) })  -\frac{1}{4}( x^2-(x-2\mathfrak{m})^2)  \;. 
\end{align} 
Extremizing this function, we can solve for $x$ as (with $X \coloneqq e^x,   M \coloneqq e^{\mathfrak{m}}$):
\begin{align}
X = \frac{M^2(M^2-1)}{1-M-(1+L)M^2} \;,
\end{align}
and by eliminating $x$ one obtains
the relation  between the longitude $\mathfrak{l}$ and the meridian $\mathfrak{m}$ (with $L \coloneqq e^l$):
\begin{align}
L+ \frac{1}{L}+ 2 - \frac{1}{M^2} + \frac{1}{M} + M - M^2 = 0 \;.
\end{align}
The latter equation is nothing but (the non-reducible part of\footnote{The A-polynomial in general contains a factor $L-1$ for the reducible flat connection. The expression here is for the irreducible flat connection.}) the A-polynomial \cite{CooperApolynomial} of the figure eight knot complement.

To this point the analysis is completely parallel to the non-supersymmetric case \cite{Terashima:2011xe}.
However, there is a difference in the next-order correction, which can be computed
by evaluating the Gaussian fluctuations around the saddle point.
For the non-supersymmetric case, the result of the computation is
\begin{align}
\widetilde{\text{Tr}}_{\textrm{non-SUSY}}(\mathsf{LR}) 
&= e^{\frac{1}{\hbar} S_0 +  
\log\hbar + \frac{1}{2}\ln S_1} (1 + O(\hbar)) \;,
\end{align}
with the one-loop part $S_1$ gives the Reidemeister torsion \cite{Porti}
\begin{align}
S_1=\frac{M^2}{(M^2-3M+1) (M^2+M+1)} \;.
\end{align}

In the supersymmetric case, there is a new contribution
\begin{align}
\widetilde{\text{Tr}}_{\rm R/NS} (\mathsf{LR}) 
&= \exp\left[\frac{1}{\mathcolor{red}{2}\hbar} S_0 +  
\log(2\hbar) + \frac{1}{2} \ln S_1 \mathcolor{red}{\pm \frac{1}{4} \ln \tilde{S}_1}\right](1 + O(\hbar)) \;,
\end{align}
where the spin structure dependent piece is given by 
\begin{align}
\tilde{S}_1=S_1\big|_{\rm saddle point}=\frac{(\sqrt{X^\star}+ i M) (1-i\sqrt{X^\star})} {(\sqrt{X^\star}- i M) (1+i \sqrt{X^\star})}\;,
\end{align}
with the saddle point expression
\begin{align}
X^\star = \frac{1}{2} \left(-1 + M - M^2 - \sqrt{1 - 2 M - M^2 - 2 M^3 + M^4}\right) \;.
\end{align}

We have seen that the Reidemeister torsion is modified in supersymmetric cases
as
\begin{align}
S_1\to S_1 \mathcolor{red}{(\tilde{S}_1^{\pm 1/2}) }\;,
\end{align}
depending on the choice of the spin structure.
The combination on the right hand side should be regarded as a supersymmetric version of the Reidemeister/Ray-Singer torsion (depending on a
spin structure of the 3-manifold), and it would be interesting to further study this torsion.\footnote{In physics language the Reidemeister torsion is 
the one-loop piece of the 3d Chern-Simons theory \cite{Witten:1988hf} (see Sec.~\ref{sec:3d3d} for connections with the $\OSp(1|2)$ Chern-Simons theory).
The torsion for a supergroup Chern-Simons theory was discussed e.g.\ in \cite{Mikhaylov:2015nsa}.}

It is straightforward to explicitly calculate the higher orders terms in the expansion
with the Feynman diagram techniques, see e.g.\ \cite{Dimofte:2009yn,Dimofte:2012qj,Gang:2015wya}.

\subsection{\texorpdfstring{General $\varphi$}{General Varphi}}

Comparison between the 
integral expressions for the partition function for $\varphi=LR$
for supersymmetric \eqref{susy_LR} and 
non-supersymmetric \eqref{nonsusy_LR} case 
makes the structure rather manifest.
Namely, we need to (1) divide the quadratic Gaussian factor by a factor of $2$
and (2) replace the quantum dilogarithm $e_b$ by their supersymmetric counterparts $\eR$ or $\eNS$, depending on the 
choice of the spin structure.
We can therefore straightforwardly adopt the results of \cite{Terashima:2011xe}
to a general element $\varphi$ of the mapping class group,
and discuss invariants of spin 3-manifolds, where the 3-manifolds in question are complements of the so-called fibered knots.
Note that a general element of $\SL(2, \mathbb{Z})$ can be written as $\varphi=L^{n_1} R^{n_2} L^{n_3} R^{n_4}\dots $
where $n_1, n_2, \dots$ are integers.
As we discuss in appendix \ref{AppendixB}, the number of spin structures of the once-punctured torus kept fixed under a general element of the $SL(2, \mathbb{Z})$ is either one, two or four.
This means that in general there are two, four or eight spin structures on 3-manifolds. As commented already, however,
we have only two different partition functions given by the choices of $\Tr_{\rm R/NS}(\hvarphi)$.

Let us further illustrate this point with the example of $\varphi=L^2R$. 
The resulting 3-manifold is listed as m009 in the SnapPea census \cite{SnapPy}.
In this example, two spin structures of the 2-manifold are fixed, type $(1,1)$ and type $(1,0)$.
We have four spin structures, nevertheless we only have two partition functions again, given by $\Tr_{\rm R/NS}(\mathsf{L L R})$.

In the non-supersymmetric case, the trace is computed as \cite{Terashima:2011xe}
\begin{align}
&\Tr_{\textrm{non-susy}}(\mathsf{LLR})=\int dy_1dx_2 dx_3  \,\,
 e^{\frac{1}{\hbar} V_{\rm quad}(y_1,x_2,x_3,h)}
e_b\left(-\frac{y_1}{2\pi b} \right) 
e_b\left(\frac{x_2}{2\pi b} \right)^{-1} 
e_b\left(\frac{x_3}{2\pi b} \right)^{-1} \;,
\end{align}
with the quadratic piece $V_{\rm quad}$ given by
\begin{align}
&V_{\rm quad}(y_1,x_2,x_3,h) = +\frac{1}{4}x_3^2+
\frac{1}{4}y_1^2-
\frac{1}{4}(x_3-x_2)^2+
\frac{1}{2}(x_3y_1-y_1x_2)+
\frac{1}{2}h(x_2-y_1) \;.
\end{align}

The supersymmetric counterparts are computed to be
\begin{align}
&\Tr_{\textcolor{red}{\rm R/NS}}(\mathsf{LLR})=\int dy_1dx_2 dx_3  \,\,
e^{\frac{1}{\textcolor{red}{2}\hbar} V_{\rm quad}}
e_{\textcolor{red}{\R/\NS}}\left(-\frac{y_1}{2\pi b} \right) 
e_{\textcolor{red}{\R/\NS}}\left(\frac{x_2}{2\pi b} \right)^{-1} 
e_{\textcolor{red}{\R/\NS}}\left(\frac{x_3}{2\pi b} \right)^{-1} \;.
\end{align}
The classical limit $\hbar\to 0$ is given by
\begin{align}
\Tr_{\rm R/NS}(\mathsf{LLR})= \int dy_1dx_2dx_3 \,
e^{ \frac{1}{2\hbar}V_{0} (y_1, x_2, x_3,h)\pm \frac{1}{4} V_{1} (y_1, x_2, x_3)} \left(1+O(\hbar)\right) \;,
\end{align}
where $V_0$ is the same expression
as in the non-supersymmetric case, found in \cite{Terashima:2011xe}:
\begin{align}
&V_{0}(y_1,x_2,x_3,h) = 
\Li_2(-e^{-y_1})-\Li_2(-e^{x_2})-\Li_2(-e^{x_3} )\nonumber\\
&\qquad +\frac{1}{4}x_3^2+
\frac{1}{4}y_1^2-
\frac{1}{4}(x_3-x_2)^2+
\frac{1}{2}(x_3y_1-y_1x_2)+
\frac{1}{2}h(x_2-y_1) \;, 
\end{align}
and $V_1$ is the new contribution for the supersymmetric case:
\begin{align}
&V_1(y_1, x_2, x_3)=\ln \frac{\mathcal{L}(-y_1)}{\mathcal{L}(x_2)\mathcal{L}(x_3)} \;.
\end{align}

The saddle point equations are the same as in the non-supersymmetric case.
For the value $h=0$ (corresponding to the complete hyperbolic structure of the 3-manifold),
and we get 
\begin{align*}
&e^{y_1}=e^{x_2}={(-1 + i\sqrt{7})/4} \;,	&&e^{x_3} = (-3 + i \sqrt{7})/8 \;,
\end{align*}
and the extremal value $S_0$ of the potential $V_0$ reproduces the
complexified volume of the 3-manifold.

We can also keep $h$ generic. After redefinition
into longitude and meridian variables
\begin{align*}
y_1 &= x_2-2 \mathfrak{m}\;, \quad
h = \mathfrak{l}-i\pi  \;,
\end{align*}
and the change of variables
$L = e^{\mathfrak{l}}, M=e^{\mathfrak{m}},  X = e^x $
we can solve the saddle point equation as
\begin{align} \label{saddle_X2X3}
 \begin{split}
X_2^\star &= \frac{1}{4} (-1 + M - M^2 - \sqrt{1 - 2 M - 5 M^2 - 2 M^3 + M^4}) \;,\\
X_3^\star &= \frac{(-1 - M - M^2 - \sqrt{1 - 2 M - 5 M^2 - 2 M^3 + M^4})}{
	2 (1 + 2 M + M^2)} \;,
 \end{split}
\end{align} 
and by elimination we reproduce the non-reducible part of the A-polynomial
\begin{align}
A(L,M)=L^{-1} + L M +  2 - 1/M + 2 M - M^2=0 \;.
\end{align}

We can also compute the subleading correction contributing to the Reidemeister torsion:
\begin{align}
\widetilde{\text{Tr}}_{\R/\NS}(\mathsf{LLR}) 
&= \exp\left[\frac{1}{\mathcolor{red}{2}\hbar} S_0 +  
\log(2\hbar) + \frac{1}{2}\ln S_1 \pm \mathcolor{red}{\frac{1}{4}\ln \tilde{S}_1}\right] (1 + O(\hbar)) \;.
\end{align}
Here $S_1$ is the one-loop determinant for the non-supersymmetric case, which reads
\begin{align}
S_1=  \frac{M^{2}}{1+M (1 + M) (-2 + (-3 + M)M )} \;,
\end{align}
and this coincides with the known expression for the Reidemeister torsion \cite{Porti}.
For supersymmetric case, we have a new contribution 
\begin{align}
\tilde{S}_1(M)=\frac{(\sqrt{X_2^\star}+iM )}{(\sqrt{X_2^\star}-iM)}
\frac{(1-i \sqrt{X_2^\star})}{(1+i \sqrt{X_2^\star})}
\frac{(1-i \sqrt{X_3^\star})}{(1+i \sqrt{X_3^\star})} \;,
\end{align}
where $X_2^\star, X_3^\star$ are the critical points presented in \eqref{saddle_X2X3}.

\section{Relation with Chern-Simons Theories}\label{sec:CS}

\subsection{\texorpdfstring{$\OSp(1|2)$ Chern-Simons Theory}{OSp(1|2) Chern-Simons Theory}}

Since the super \Teichmuller space is a subspace of the moduli space of flat $\OSp(1|2)$ connections, 
it is natural to imagine that the spin TQFT 
associated with the 3-manifold
should also be related to another quantization of the moduli space of the flat $\OSp(1|2)$ connections,
namely the 3-dimensional Chern-Simons theory with gauge group $\OSp(1|2)$.

One should quickly add that it is far from clear if this reasoning indeed works. The super \Teichmuller space is only a subspace 
of full space of flat $\OSp(1|2)$ connections, however in the formulation of 
TQFT one needs to sum over all the possible states when factorizing the geometry,
and hence it is not clear how one can restrict to a subspace consistently. 

A similar problem was discussed for the non-supersymmetric \Teichmuller theory \cite{Mikhaylov:2017ngi}. In this paper, it was 
explained that the \Teichmuller TQFT arises from the 
complex Chern-Simons theory on a particular integration contour
specified by the singular Nahm pole boundary condition.
While we do not work out all the details,
we expect that a similar reasoning will 
guarantee that our super \Teichmuller spin TQFT can be identified with 
the complex $\OSp(1|2)$ Chern-Simons spin TQFT on a certain integration contour. 
Note that the supergroup
$\OSp(1|2)$ Chern-Simons theory can be thought of as a theory of 
gauge fields with fermions, and hence depends on the 
choice of the spin structure of the 3-manifold and is a spin TQFT.

\subsection{\texorpdfstring{Duality to $\SU(2)$ Chern-Simons Theory}{Duality to SU(2) Chern-Simons Theory}}

We can now appeal to the duality discussed in \cite{Mikhaylov:2014aoa} (see also \cite{MR1188811,MR3704249}).
By analyzing $\SL(2,\mathbb{Z})$-duality of a topologically-twisted 4-dimensional supersymmetric Yang-Mills theory \cite{Kapustin:2006pk},
one obtains the duality between analytic continuations of 
3-dimensional Chern-Simons theories with different gauge groups:
$\OSp(2m+1|2n)$ and $\OSp(2n+1|2m)$.
Their duality works when one identifies the coupling constants of the two theories
up to a sign flip:
\begin{align}
q_{\OSp(2m+1|2n)}= -q_{\OSp(2n+1|2m)} \;.
\end{align}

For our purpose, we can consider the special case
of $m=0, n=1$, which gives a duality between
$\OSp(1|2)$ theory and $\SO(3)$ theory.
The statement is that the 
analytic continuation of the 
$\OSp(1|2)$ theory coincides with that of the
$\SO(3)$ theory.

At the level of Lie algebras $\mathfrak{so}(3)$
coincides with $\mathfrak{su}(2)$, whose complexification 
gives $\mathfrak{sl}(2, \mathbb{C})$.
Since the real slice of $\mathfrak{sl}(2, \mathbb{C})$
is $\mathfrak{sl}(2, \mathbb{R})$, and since this is relevant for 
\Teichmuller theory, one expects
that the resulting theory is essentially the (non-supersymmetric) quantum
\Teichmuller TQFT. This naively seems to be in tension with the fact that 
the other side of the duality, the $\OSp(1|2)$ theory,
is a spin TQFT, not a TQFT.

The apparent tension is resolved by the subtle difference between $\SO(3)$ and $\SU(2)$ gauge groups.\footnote{The differences between 
$\SO(3)$ and $\SU(2)$ gauge group for Chern-Simons theory 
plays crucial roles in the formulation of the closed 3-manifold version of the volume conjecture \cite{Chen:2015wfa,Gang:2017cwq},
which involves specifications of integration contours in Chern-Simons theory \cite{Witten:2010cx}.}
Recall that the gauge-invariance of the 
Chern-Simons action (for a gauge group $G$)
\begin{align}
S_{\rm CS}=\frac{k}{4\pi}  \int_M \textrm{Tr}_G\left( A\wedge dA + \frac{2}{3} A\wedge A\wedge A \right) \;,
\end{align}
is guaranteed by the quantization of the integral of the characteristic class
\begin{align}
N \coloneqq \frac{1}{8\pi^2}  \int_{M_4} \textrm{Tr}_{G} \left( F\wedge F \right) \;,
\end{align}
where $M_4$ is a closed four-manifold: the level $k$ is an integer if $N\in \mathbb{Z}$.
Now, when the quantization condition is $N \in \mathbb{Z}$
for $G=\SU(2)$, the corresponding quantization condition is $N \in \mathbb{Z}/4$
for $G=\SO(3)$. This means that  the $\SO(3)$ Chern-Simons theory with the smallest level 
($k_{\rm SO(3)}=1$) corresponds to the level $k_{\rm \SU(2)}=4$ $\SU(2)$ Chern-Simons theory.

The situation is different if 
we further assume that both $M$ and $M_4$ are spin manifolds. In this case,
the quantization condition for $G=\SO(3)$ is now $N \in \mathbb{Z}/2$,
so that the minimal choice of the level for the $\SO(3)$ theory corresponds to 
level $k_{\rm \SU(2)}=2$ in the $\SU(2)$ Chern-Simons theory. Of course, this means that the 
$\SO(3)$ theory is now a spin TQFT, which is what we expect when we discuss the duality with
$\OSp(1|2)$ Chern-Simons theory.

Summarizing, we find that $\OSp(1|2)$ spin TQFT
should be identified with the \Teichmuller TQFT at level $2$,
under the identification $q_{\OSp(1|2)}= -q_\text{Teichm{\" u}ller}$.

The connection with the level-$2$ $\SU(2)$ Chern-Simons theory can be 
worked out more explicitly from concrete expressions.
In the level-$2$ Chern-Simons theory studied in \cite{Dimofte:2014zga},
the basic building blocks are the ``level $2$ version'' of the quantum dilogarithm function,
which are given by\footnote{In the notation closer to \cite{Dimofte:2014zga}, one has
\begin{align}
\begin{split}
\mathcal{Z}_\ub^{k=2}(x, 0)&=\prod_{\gamma, \delta=0,1, \gamma-\delta \equiv 0 (\mathrm{mod} 2)} \mathcal{Z}_\ub\left(\frac{x+i \ub\gamma+i \ub^{-1} \delta}{2}\right) 
=\mathcal{Z}_\ub\left(\frac{x}{2}\right) \mathcal{Z}_\ub\left(\frac{x+i \ub+i \ub^{-1}}{2}\right) \;,\\
\mathcal{Z}_\ub^{k=2}(x, 1)&=\prod_{\gamma, \delta=0,1, \gamma-\delta \equiv 1 (\mathrm{mod} 2)} \mathcal{Z}_\ub\left(\frac{x+i \ub\gamma+i \ub^{-1} \delta}{2}\right) 
=\mathcal{Z}_\ub\left(\frac{x+i b}{2}\right) \mathcal{Z}_\ub\left(\frac{x+i \ub^{-1}}{2}\right) \;.
\end{split}
\label{eq:psi_k}
\end{align}
This is converted to by \eqref{eb_k} by the relation $\mathcal{Z}_\ub(x)=e_\ub(-(x+i(\ub+ \ub^{-1})/2))$.}
\begin{align}\label{eb_k}
\begin{split}
\mathcal{Z}_\ub^{k=2}(x, 0)&\coloneqq 
e_\ub\left(\frac{-x+i(\ub+\ub^{-1})\slash2}{2}\right)e_\ub\left(\frac{-x-i(\ub+\ub^{-1})\slash2}{2}\right) =e_\NS(-x) \;,\\
\mathcal{Z}_\ub^{k=2}(x, 1)&\coloneqq
e_\ub\left(\frac{-x+i(\ub-\ub^{-1})\slash2}{2}\right)e_\ub\left(\frac{-x-i(\ub-\ub^{-1})\slash2}{2}\right)=e_\R(-x) \;.
\end{split}
\end{align}
These are (up to a sign) nothing but the definitions of the ``NS'' and ``Ramond'' 
quantum dilogarithms introduced earlier in \eqref{e_SUSY}. 

One can also find the shift $q_{\OSp(1|2)}= -q_\text{Teichm{\" u}ller}$.
Recall that in our notation we had $q_{\OSp(1|2)}=e^{i\pi \ub^2}$.
This should be compared with the definition of the 
$q$-parameter in \cite{Dimofte:2014zga}:
\begin{align}\label{qTh}
q_{\text{\Teichmuller}} =\exp\left(\frac{2\pi i}{2}(\ub^2+1) \right) \;,
\end{align}
which indeed satisfies the sign shift $q_{\OSp(1|2)}= -q_\text{Teichm{\" u}ller}$.
Note that the combination \eqref{qTh} can be derived from 
supersymmetric localization of 5d $\mathcal{N}=2$ theory \cite{Cordova:2016cmu}.

Note that the discussion of the level $2$ Chern-Simons theory in \cite{Dimofte:2014zga}
does not mention the spin structure, and the theory there is meant to be a TQFT, not a spin TQFT.
This is not necessarily a contradiction since given a spin TQFT
one can define a topological partition function by summing the spin TQFT partition function
over possible spin structures. We have already seen a version of this when we discussed the trace 
in the super \Teichmuller theory, which we now know to be a 
sum of two integral expressions for two different spin structures:
\be
\Tr(\hvarphi)=\Tr_{\rm NS}(\hvarphi)+ \Tr_{\rm R}(\hvarphi) \;,
\ee
where the right hand side is a sum over the $\mathbb{Z}_2$ choices 
along the base ($S^1$) direction of the 
mapping torus (the spin structure is still chosen and fixed along the fiber ($\Sigma$) direction).

By turning the argument around, one could expect that the 
partition functions of analytically-continued $\OSp(1|2)$ Chern-Simons theory on general 3-manifolds
 can be obtained from those of the level $2$ analytically-continued $\SU(2)$ Chern-Simons theory
 by applying suitable projection operators.

\section{Chain of Connections: Super 3d--3d Correspondence}\label{sec:3d3d}

In this section let us comment on connections with various topics.
We keep our discussion short and we will be content here with sketching the main ideas. Each of the topics deserves a serious study, which we leave for future work.

\paragraph{3d $\mathcal{N}=2$ Theories on $\mathbb{RP}^3$ and Super 3d--3d correspondence:}
We have seen that the super \Teichmuller spin TQFT is eventually related to the
level $2$ complex Chern-Simons theory.

One of the motivations for the level-$k$ Chern-Simons theory \cite{Dimofte:2014zga}
was to consider the 3d--3d correspondence \cite{Terashima:2011qi,Dimofte:2011ju}.\footnote{See \cite{Yagi:2013fda,Lee:2013ida,Cordova:2013cea} for direct derivations from supersymmetric localization of the five-dimensional $\mathcal{N}=2$ theory.} Here for a 3-manifold $M$ there is a natural 3d $\mathcal{N}=2$ theory $\mathcal{T}[M]$,
whose supersymmetric partition function on the lens space $L(k,1)\simeq S^3/\mathbb{Z}_k$ coincides with the Chern-Simons partition function on $M$. 
In fact, historically the expression for the ``level-$k$'' quantum dilogarithm
\eqref{eq:psi_k} was derived first in the context of supersymmetric localization of 3d $\mathcal{N}=2$ theories on the lens space  \cite{Gang:2019juz,Benini:2011nc},
which was then used as the building block for the level-$k$ discussion of \cite{Dimofte:2014zga}.

In this context, one can track the two choices of  the quantum dilogarithm function into the choice of discrete 
$\mathbb{Z}_2$ holonomies along the $S^1$-circle of $\mathbb{S}^3/\mathbb{Z}_2$;
periodic and anti-periodic boundary conditions for fermions.

The comment of the previous paragraph suggests that there should be a refinement of the level-$2$ Chern-Simons theory
corresponding to a specific choice of the spin structure. In other words, the expectation is that 
\begin{align}
\begin{split}
\textrm{($\mathcal{N}=2$ theories $\mathcal{T}[M]$ on $\mathbb{RP}^3$, with a fixed boundary condition for fermions)}
\\
\leftrightarrow \textrm{(Complex $\OSp(1|2)$ Chern-Simons theory on a spin 3-manifold $M$)} \;. 
\end{split}
\end{align}

There is another path to arrive at the same conclusion,
using a chain of connections invoked for the non-supersymmetric case \cite{Terashima:2011qi,Cordova:2016cmu}.
Let us comment on this now.

\paragraph{Super Liouville Theory:}

The first piece of the chain is
the connection with quantum super \Teichmuller theory 
and the quantum super Liouville theory. 
While these two theories are apriori 
different theories quantum-mechanically, we conjecture
that the two quantizations are actually equivalent.
The bosonic analog of this statement,
that the quantum \Teichmuller theory 
coincides with quantum Liouville theory,
was conjectured in \cite{Verlinde:1989ua}
and was studied in detail in \cite{Teschner:2003at,Teschner:2005bz,Teschner:2010je}
(see also pioneering works \cite{Faddeev:2000if,Faddeev:2002ms}).

\paragraph{Four-dimensional $\mathcal{N}=2$ Theory:}

We can next look at the connection between Liouville theory
and 4-dimensional $\mathcal{N}=2$ theory {\`a} la
Alday, Gaiotto and Tachikawa \cite{Alday:2009aq}.
While the original proposal referred to the 
non-supersymmetric Liouville theory,
an extension to super Liouville theory was later discussed in \cite{Belavin:2011pp,Bonelli:2011jx,Bonelli:2011kv}.\footnote{While the proposal of \cite{Belavin:2011pp} involved a decoupled coset, it is argued in \cite{Cordova:2016cmu} that
this is taken into account by complexifying the Liouville theory.}
Namely, the conformal blocks of $\mathcal{N}=1$ super Liouville theory
were identified with the instanton partition functions of 4-dimensional $\mathcal{N}=2$ theories 
on $\mathbb{R}^4/\mathbb{Z}_2$. 
One expects that this is the ``half'' of the 
supersymmetric partition function on $\mathbb{S}^4/\mathbb{Z}_2$, generalizing the results of \cite{Pestun:2007rz,Hama:2012bg}.

\paragraph{3d $\mathcal{N}=2$ Theory:}

In the proposal of \cite{Terashima:2011qi},
the 3d $\mathcal{N}=2$ theories associated with mapping tori
are identified with duality domain wall theories
inside 4d $\mathcal{N}=2$ theories.
In the context of the supersymmetric localization,
the domain wall occupies the equator $\mathbb{S}^3$
inside the $\mathbb{S}^4$.
In our context, $\mathbb{S}^4$ is replaced by 
$\mathbb{RP}^4=\mathbb{S}^4/\mathbb{Z}_2$, and hence the equator 
$\mathbb{S}^3$ should be replaced by $\mathbb{RP}^3=\mathbb{S}^3/\mathbb{Z}_2$. 

By combining all the three arguments above, we have thus arrived at the same conclusion as
before: super \Teichmuller TQFT should be identified with the 
3d $\mathcal{N}=2$ theories on $\mathbb{RP}^3$.

\paragraph{M5-branes on $\mathbb{RP}^3\times M$:}

We have seen that super \Teichmuller TQFT is related with a
number of different topics in physics and mathematics.
The ultimate reason for these statements is that
all these theories arise from the compactification of the 
6d $\mathcal{N}=(2,0)$ theory on $\mathbb{RP}^3$:
\begin{align}
\textrm{(6d $\mathcal{N}=(2,0)$ theory  on $\mathbb{RP}^3\times M$)} \leadsto
\textrm{(super \Teichmuller spin TQFT on  $M$)} \;.
\end{align}
By compactifying the 6d theory along the Hopf fiber direction of $\mathbb{RP}^3=S^3/\mathbb{Z}_2$
this should have a direct derivation from supersymmetric localization of five-dimensional $\mathcal{N}=2$ theory
along the lines of \cite{Yagi:2013fda,Lee:2013ida,Cordova:2013cea}.

\section{Future Directions}\label{sec:summary}

Let us end this paper by listing several open problems.

\begin{itemize}[wide, labelwidth=!, labelindent=10pt]

\item One of the most important questions is to formulate the super \Teichmuller spin TQFT for a 
general spin 3-manifold. For this purpose, one needs combinatorial 3d spin structures on 3d ideal tetrahedra. While  combinatorial spin structures on 3-manifolds
have been discussed in \cite{MR3180826,MR3784005,Gaiotto:2015zta}, in the literature there seems to be
no known Kasteleyn-type combinatorial spin structure for 3-manifolds convenient for our purposes---such a combinatorial spin structure for the 3-manifold
should reduce to the Kasteleyn orientation on the boundary 2-manifold.
It should in principle be possible to ``uplift'' our 2d Kasteleyn orientations to 
3d Kasteleyn orientations. For example a flip in the 2-dimensional surface can be uplifted to a 3-dimensional tetrahedron,
and the 2d Kasteleyn orientation can be uplifted into allowed orientations of 
3-dimensional tetrahedron.  Moreover, the pentagon relation in the 2-dimensional case can be interpreted as
the 3-dimensional $2-3$ Pachner move, now equipped with 3d Kasteleyn orientations.

\item There is a natural generalization the super \Teichmuller theory,
where $\OSp(1|2)$ flat connections are replaced by $\OSp(N|2)$ flat connections.
For the special case of $N=2$, this is the 
$\mathcal{N}=2$ super \Teichmuller theory,
whose classical theory was discussed in \cite{Ip:2016ojn}.
We expect that the resulting partition function will depend on the choice of the ``para-spin'' structure,
where the role of $H^1(M, \mathbb{Z}_2)$ is played by $H^1(M, \mathbb{Z}_N)$ (this is related with the spin$^c$ structure).
In the context of Liouville theory (see Sec.~\ref{sec:3d3d}), $N>1$ counterparts of our theories are
the para-Liouville theories \cite{Argyres:1990aq,LeClair:1992xi}.
In \cite{Belavin:2011tb,Nishioka:2011jk} (see also \cite{Belavin:2011pp,Bonelli:2011jx,Bonelli:2011kv})
the connection between the para-Liouville theory and the 
instanton counting on $\mathbb{R}^4/\mathbb{Z}_N$ was discussed.
By the similar logic as before, this should be related with
the 3d $\mathcal{N}=2$ theory on
$S^3/\mathbb{Z}_N$, which 
in turn can be related with complex Chern-Simons theory with level $N$.\footnote{One can further consider further generalization to $M$-th para-Toda theories,
which correspond to complex Chern-Simons theories with $\SU(M)$ gauge groups at level $N$.}

\item The semiclassical analysis of this paper can be regarded as the unity limit $b\to 0, q\to 1 $ of the 
quantum invariants. It would be interesting to explore expansion around more general rational points,
see \cite{Garoufalidis:2014ifa,Ip:2014pva} for related discussion.

\item We  can include supersymmetric defects to the super 3d--3d correspondence discussed in Sec.~\ref{sec:3d3d}.
We can either consider co-dimension $2$ defects of co-dimension $4$ defects in the 6d theory 
(see the analysis for the non-supersymmetric cases in e.g.\ \cite{Coman:2015lna,Frenkel:2015rda,Gang:2015bwa,Gang:2015wya}).
For example, an insertion of a co-dimension $4$ defect is represented by an insertion of a Wilson line operator $\hat{W}$ 
inside the trace, so that we have an expression of the form $\Tr(\hvarphi \hat{W})$ \cite{Gang:2015bwa}.

\item Instead of a mapping torus with a
non-trivial twist, we can consider a mapping torus without a twist, namely the direct product
$\Sigma\times S^1$. We can then reduce to the 
2-dimensional theory.
The resulting 2-dimensional theory is the BF theory \cite{Blau:1993tv}, this time
associated with the supersymmetric version \cite{Montano:1990ru,Chamseddine:1991fg,Cangemi:1993mj} of the Jackiw-Teitelboim gravity \cite{Teitelboim:1983ux,Jackiw:1984je}.
This theory has recently been studied in connection
with the supersymmetric extension of the SYK model \cite{Stanford:2019vob}.
This suggests that some of the techniques of this paper could have applications there.
\end{itemize}

\section*{Acknowledgements}
We would like to thank IHES (Summer School ``Supersymmetric Localization and Exact Result''), DESY and MPIM Bonn for hospitality. We would like to especially thank J\"{o}rg Teschner for useful discussions and comments. We would also like to thank Dongmin Gang, Anton Kapustin, Rinat Kashaev, Victor Mikhaylov and Volker Schomerus for discussions. 
The research of M.Y.~was supported in part by WPI Research Center Initiative, MEXT, Japan, and by the JSPS Grant-in-Aid for Scientific Research (No.\ 17KK0087, No.\ 19K03820 and No.\ 19H00689). The work of M.K.P. was supported by the European Research Council (advanced grant NuQFT). The work of N.A. was supported by Max Planck Institute of Mathematics (MPIM) in Bonn and the Swiss National Science Foundation (pp00p2-157571/1).

\appendix


\section{Spin Structure versus Kasteleyn Orientation}\label{appendixA}
In this appendix we summarize the relation between the Kasteleyn orientation and the spin structure. 

Let us fix a Kasteleyn orientation on the 2-manifold $\Sigma$.
Suppose we choose a closed oriented cycle $C$  on $\Sigma$: namely $C$ is a set of edges in the hexagonalization of the ideal triangulation. 
Let $K(C)$ be the number of the edges where the orientation along the path $C$ is opposed from the orientation determined by the Kasteleyn orientation.
By $l(C)$ we denote the number of left dimer sticks along the path $C$, where the dimer stick is defined as the small continuation of all edges at the vertices of the hexagon. 
 If we reach a dimer sticks along the a path $C$ on the left hand side of the path we call it a left dimer stick. 
 
\begin{figure}[htbp] 
	\centering	\includegraphics[width=0.3\textwidth]{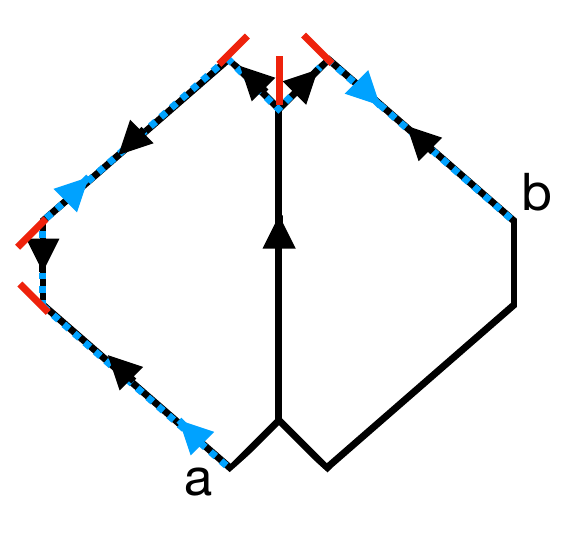}
	\caption{An example of a path $C$ (from point $a$ to $b$ on the once-punctured torus) with $K(C)=4$, $l(C)=5$.}
	\label{l(C)}
\end{figure}

 Let us explain this notation using Fig.~\ref{l(C)}. We start a blue path $C$ from point $a$ to point $b$. We have four of the edges with orientations against the Kasteleyn orientation and therefore $K(C)=4$. Along the blue path we find the left dimer sticks (shown with red color) five times and therefore $l(C)=5$.

Let us define a $\mathbb{Z}_2$ sign
$q(C)$ associated with the closed path $C$ by
\begin{equation}\label{spin}
q(C) \coloneqq 1+K(C)+l(C)  \quad (\textrm{mod} \, 2 ) \;.
\end{equation}
It turns out that this sign depends only on the spin structure (i.e.\ the equivalence class of the Kasteleyn orientations)
and on the cohomology class of the path $C$.
The signs can thus be regarded as a map from a spin structure
to an element of $H^1(\Sigma, \mathbb{Z}_2)$ (\cite[Theorem 1]{MR2410902}, see also \cite[Theorem 3.2.8]{BB}).

\bigskip 

\begin{figure}[htbp] 
	\centering
	\includegraphics[width=1\textwidth]{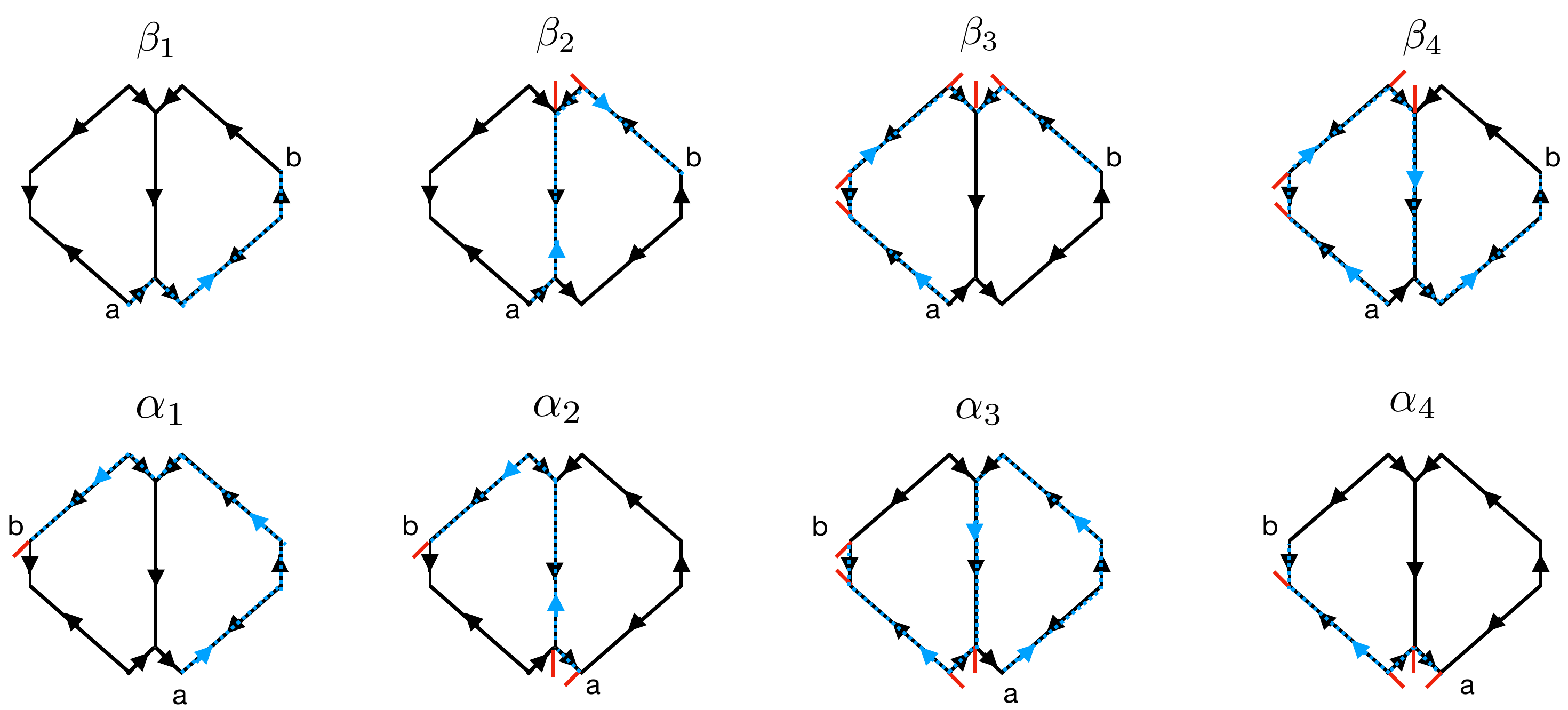}
	\caption{Four representatives for the $\alpha$ and $\beta$-cycles in the cohomology $H^1(\Sigma,\mathbb{Z})$ of the once-punctured torus. Here we have fixed the Kasteleyn orientation corresponding to case $(0,0)$ of Fig.~\ref{spins}. The resulting $\mathbb{Z}_2\oplus \mathbb{Z}_2$ signs
	are independent of the choice of the representative.}	
	\label{c}
\end{figure}

As an example, we consider the once-punctured torus.
We have four equivalence classes of Kasteleyn orientations as in Fig.~\ref{spins}.
Let us here discuss the case $(0,0)$. To find $l(C)$ we can draw the $\alpha$ and $\beta$-circle in the hexagonalization
as in Fig.~\ref{c}. We show four different 
representative paths for the cohomology class, both for the $\alpha$ and the $\beta$-cycle.
The point $a$ is the starting point of the path $C$ and the red dimers are those dimers which are on the left side of the path $C$.
The result for different choice of path $C$ is the same, as long as they represent the same cohomology class in $H^1(\Sigma, \mathbb{Z})$.

In our example of type $(0,0)$ of Fig.~\ref{spins}, we find \\
\begin{align}
\begin{array}{c||cccc||cccc}
& \alpha_1 &\alpha_2 &\alpha_3 &\alpha_4 & \beta_1 &\beta_2 &\beta_3 &\beta_4 \\
\hline
\hline
K(C) &1  &3  & 4 & 3&2  &3  & 3& 3\\
\hline
l(C) & 0 & 2 & 5 & 4&1  &2  & 4 & 4\\
\hline
q(C) [\mathrm{mod} \, 2]  & 0 & 0 & 0 & 0&0 &0  & 0 & 0\\
\multicolumn{1}{c}{\raisebox{.8\normalbaselineskip}[0pt][0pt]{$$}} & \multicolumn{4}{c}{\raisebox{.8\normalbaselineskip}[0pt][0pt]{$\underbrace{\phantom{12345678910}}_{q(\alpha)}$}} & 
\multicolumn{4}{c}{\raisebox{.8\normalbaselineskip}[0pt][0pt]{$\underbrace{\phantom{12345678910}}_{q(\beta)}$}} 
\end{array}
\end{align}\\
and therefore, $q(\alpha)\equiv 0, q(\beta)\equiv 0 \,\, (\mathrm{mod} \, 2)$ irrespective of the choice of the representative for the cohomology class, as expected.
One can repeat this exercise for the other three Kasteleyn orientations in Fig.~\ref{spins}.

\newpage
\section{Spin Mapping Class Group}\label{AppendixB}

In this appendix we study the number of spin structures kept fixed under a general element of the $SL(2, \mathbb{Z})$.

A general element of $\SL(2, \mathbb{Z})$ can be written as $\varphi=L^{n_1} R^{n_2} L^{n_3} R^{n_4}\dots $
where $n_1, n_2, \dots$ are integers. Since $L^2$ and $R^2$ preserves the spin structure, for our purposes we can consider the integers $n_1, n_2, \dots$ 
modulo $2$, so that we have elements of the form $LRLR \dots$ or $RLRL \dots$. Moreover,
since $(LR)^3$ and $(RL)^3$ preserves the spin structure, the discussion reduces to the following six cases (notice for example $LRLRL$ and $R$ act on the spin structures in the same way):
\begin{align}
\mathbb{I} \;, \quad L    \;, \quad RL \;, \quad LRL \;, \quad LR \;,  \quad R\;.
\end{align}
For each case, the action on the spin structure on each of the four spin structures can be worked out as
\begin{align}
\begin{array}{c|cccccc}
&\mathbb{I} &  L & RL & LRL & LR & R \\
\hline
(1,1) & \textcolor{red}{(1,1)} &  \textcolor{red}{(1,1)}  & \textcolor{red}{(1,1)} & \textcolor{red}{(1,1)}  & \textcolor{red}{(1,1)} & \textcolor{red}{(1,1)} \\
(1,0) &\textcolor{red}{(1,0)}& (0,0)& (0,1) &(0,1) &(0,0)& \textcolor{red}{ (1,0)} \\
(0,1) &\textcolor{red}{(0,1)}&  \textcolor{red}{(0,1)} & (0,0) & (1,0) & (1,0) & (0,0) \\
(0,0) & \textcolor{red}{(0,0)}&  (1,0) & (1,0) &  \textcolor{red}{(0,0) }&( 0,1) & (0,1) \\
\end{array} \;,
\end{align}
where those spin structures fixed under the mapping class group action are highlighted in red.
This means that in general a mapping class group element preserves either one, two or four spin structures.


\bibliography{superCS}
\bibliographystyle{nb}
\end{document}